\documentclass[lettersize,journal]{IEEEtran}
\usepackage{amsmath,amsfonts}
\usepackage{algorithmic}
\usepackage{algorithm}
\usepackage{array}
\usepackage{textcomp}
\usepackage{stfloats}
\usepackage{url}
\usepackage{verbatim}
\usepackage{graphicx}
\usepackage{cite,comment}
\usepackage{amssymb}
\usepackage{subfigure}
\usepackage{makecell}
\usepackage{xcolor}
\usepackage{xcolor}
\usepackage{xpatch}
\makeatletter
\def\changeBibColor#1{%
	\in@{#1}{}
	\ifin@\color{blue}\else\normalcolor\fi
}
\xpatchcmd\@bibitem
{\item}
{\changeBibColor{#1}\item}
{}{}
\makeatother
\newcommand{\mv}[1]{\mbox{\boldmath{$ #1 $}}}
\newtheorem{proposition}{Proposition}
\newtheorem{definition}{Definition}
\newtheorem{remark}{Remark}
\makeatletter
\renewcommand{\maketag@@@}[1]{\hbox{\m@th\normalsize\normalfont#1}}
\makeatother
\renewcommand{\baselinestretch}{0.97}
\begin{document}

\title{STAR-RIS-Enabled Multi-Path Beam Routing with Passive Beam Splitting}

\author{Bonan An,~\IEEEmembership{Student Member,~IEEE}, Weidong Mei,~\IEEEmembership{Member,~IEEE}, Yuanwei Liu,~\IEEEmembership{Fellow,~IEEE}, Dong Wang and Zhi Chen,~\IEEEmembership{Senior Member,~IEEE}
{
\thanks{Part of this paper has been presented at the IEEE Global Communications Conference (Globecom), Cape Town, South Africa, 2024.\cite{anstarris}}
\thanks{B. An is with the School of Information and Communication Engineering, University of Electronic Science and Technology of China, Chengdu 611731, China (e-mail: bonanan@outlook.com).}
\thanks{W. Mei, D. Wang and Z. Chen are with the National Key Laboratory of Wireless Communications, University of Electronic Science and Technology of China, Chengdu 611731, China (e-mail: wmei@uestc.edu.cn, DongwangUESTC@outlook.com, chenzhi@uestc.edu.cn).}
}
\thanks{Y. Liu is with the Department of Electrical and Electronic Engineering, The University of Hong Kong, Hong Kong (email: yuanwei@hku.hk).}
}
\maketitle

\begin{abstract}
	Reconfigurable intelligent surfaces (RISs) or intelligent reflecting surfaces (IRSs) can be densely deployed in the environment to create multi-reflection line-of-sight (LoS) links between base stations (BS) and users, thereby significantly enhancing the BS coverage. However, conventional reflection-only RISs can only achieve half-space reflection, which limits the LoS path diversity. In contrast, simultaneously transmitting and reflecting reconfigurable intelligent surfaces (STAR-RISs) can split incident signals into reflected and transmitted signals pointing to different half spaces simultaneously, thereby creating more LoS paths. Hence, in this paper, we study a new multi-STAR-RIS-aided communication system, where a multi-antenna BS transmits to multiple single-antenna users by exploiting the signal beam routing over a set of cascaded LoS paths each formed by multiple STAR-RISs. To reveal essential insights, we first consider a simplified single-user case, aiming to maximize its received signal power by jointly optimizing the active beamforming at the BS, the BS’s power allocation over different paths, the number of selected beam-routing paths, the selected STAR-RISs for each path, as well as their amplitude and phase shifts for transmission/reflection. However, this problem is particularly difficult to be optimally solved as different paths may be intricately coupled at their shared STAR-RISs. To tackle this difficulty, we first derive the optimal solutions to this problem in closed-form for a given set of paths. The clique-based approach in graph theory is then applied to solve the remaining multi-path selection problem efficiently. Next, we extend the proposed clique-based method to the more general multi-user case to maximize the minimum received signal power among all users, subject to additional constraints on the disjointness of the selected paths for different users. Simulation results show that our proposed STAR-RIS-enabled beam routing outperforms the conventional beam routing with reflection-only RISs in both single- and multi-user cases. 
\end{abstract}


\section{Introduction}

Reconfigurable intelligent surfaces (RISs) or intelligent reflecting surfaces (IRSs) have gained significant attention as a cutting-edge technology in the sixth-generation (6G) wireless communication systems. The RIS/IRS is able to proactively reshape the signal propagation environment by dynamically adjusting the phase-shifts of its large number of passive reflecting elements. Moreover, the RIS/IRS can operate without the need of active radio-frequency (RF) chains for transmitting or receiving signals, which substantially reduces the hardware cost and energy consumption compared to traditional active transceivers and relays \cite{9326394,9424177,renzo2019smart,8910627,basar2019wireless}. These attractive benefits of RISs/IRSs have spurred considerable enthusiasm in investigating their optimal reflection or passive beamforming designs in various wireless systems in the past several years (see e.g., \cite{9326394,9424177,renzo2019smart,8910627,basar2019wireless,di2020smart,yuan2021reconfigurable,pan2022overview,zheng2022survey} and the references therein).

However, existing works on RISs/IRSs have mostly focused on the single-reflection links by one or multiple RISs but ignored the more general multi-RIS-reflection links, which can also be utilized to further improve the end-to-end channel condition. For example, in a complex environment with dense obstacles, multi-RIS reflections can create more available cascaded line-of-sight (LoS) links to bypass the obstacles between the base station (BS) and remote user locations, thereby significantly enhancing the BS’s signal coverage. Moreover, multiple RISs can cooperatively reflect the signal over each cascaded LoS path, providing pronounced cooperative passive beamforming (CPB) gain to compensate for the severe multiplicative path loss over it \cite{pathloss}. Hence, multi-IRS-reflection aided wireless communication systems have been studied in several prior works. For example, to improve the coverage range of Terahertz (THz) communications, the authors of \cite{9410457} proposed a deep reinforcement learning (DRL)-based approach to jointly optimize the CPB of multiple RISs in a given multi-RIS-reflection link. To more effectively exploit the multi-IRS-induced LoS path diversity, some existing works \cite{mei2020cooperative,mei2021multi,9829192,10589431,mei2022intelligent,10643789,zhang2022multi,9900387,9800900,10057425,9745078,9868343} have delved into a new cooperative beam routing problem. Specifically, the authors of \cite{mei2020cooperative,mei2021multi,9829192,10589431} aimed to select an optimal multi-RIS-reflection path from the BS to each user and jointly optimize the BS/RIS active/passive beamforming in each selected path, such that the end-to-end BS-user channel power gain is maximized. A more general multi-path beam routing scheme was introduced in \cite{mei2022intelligent}, where multiple paths can be selected for each user by exploiting the active/passive beam splitting/combining at the BS/each user. Furthermore, the authors of \cite{10643789} proposed a joint resource allocation and multi-path beam routing scheme to further improve the performance of multi-user communication systems. Unlike the above works focusing on passive RISs only that may suffer high multiplicative path loss in the multi-hop reflection, the authors of \cite{zhang2022multi} investigated a general multi-active and multi-passive-RIS enabled beam routing problem, where multiple active RISs were introduced to enhance the strength of the multi-reflection links by leveraging their amplification capability. Besides, the authors of \cite{9900387} studied a new cooperative dual-polarized RIS-aided communication system and its associated beam-routing path selection problem. On the other hand, the authors of \cite{9800900,10057425,9745078,9868343} have conducted performance analysis of various metrics of multi-RIS-reflection communication systems, e.g., ergodic capacity, average symbol error probability, outage probability and sum rate.

However, all of the above studies only focused on conventional reflection-only RISs that can only achieve $180^\circ$ half-space reflection, severely limiting the LoS path diversity. To overcome this limitation, a new type of RIS, known as simultaneously transmitting and reflecting RIS (STAR-RIS), has been introduced recently, which is able to achieve 360$^\circ$ full-space manipulation of the impinging signal. In particular, the incident signal to each STAR-RIS element from either side can be split into a reflected signal and a refracted/transmitted signal pointing to different half spaces at the same time (referred to as {\it passive beam splitting} in this paper) \cite{mu2021simultaneously,9690478,10550177}. As such, by exploiting the passive beam splitting of multiple STAR-RISs, it is anticipated that more cascaded LoS paths can be created compared to conventional reflection-only RISs. Note that single-hop STAR-RISs have been extensively studied in the literature. For example, the authors of \cite{9863732} aimed to jointly optimize the transmit and reflect beamforming of a STAR-RIS in both non-orthogonal multiple access (NOMA) and orthogonal multiple access (OMA) systems. To maximize the sum rate of multiple users, the authors of \cite{wu2021coverage} jointly optimized the placement of multiple STAR-RISs and their passive beamforming. Furthermore, the authors of \cite{9856598,10049110,10050140,9786807,9915477,10049460} analyzed different performance metrics of the STAR-RIS-aided NOMA system, e.g., outage probability, ergodic rate, throughput and sum-rate. Different from previous works assuming independent phase-shift control for transmission and reflection, the authors of \cite{9838767} developed a more practical coupled transmission and reflection phase-shift model and studied its associated power consumption minimization problem. The authors of \cite{9849460} proposed to mount the STAR-RIS on an unmanned aerial vehicle (UAV) and jointly optimized the UAV's trajectory and active beamforming, as well as the passive transmission/reflection beamforming of the STAR-RIS to maximize the sum-rate of multiple ground users. The authors of \cite{10050406} proposed a STAR-RIS-enabled integrated sensing and communications (ISAC) framework with a novel sensing-at-STAR-RIS structure. Despite the extensive prior works on single-hop STAR-RISs, there is no existing work focusing on the use of STAR-RISs for multi-hop beam routing, to the best of our knowledge.

\begin{figure}[!t]
	\centering
	\includegraphics[width=3.3in]{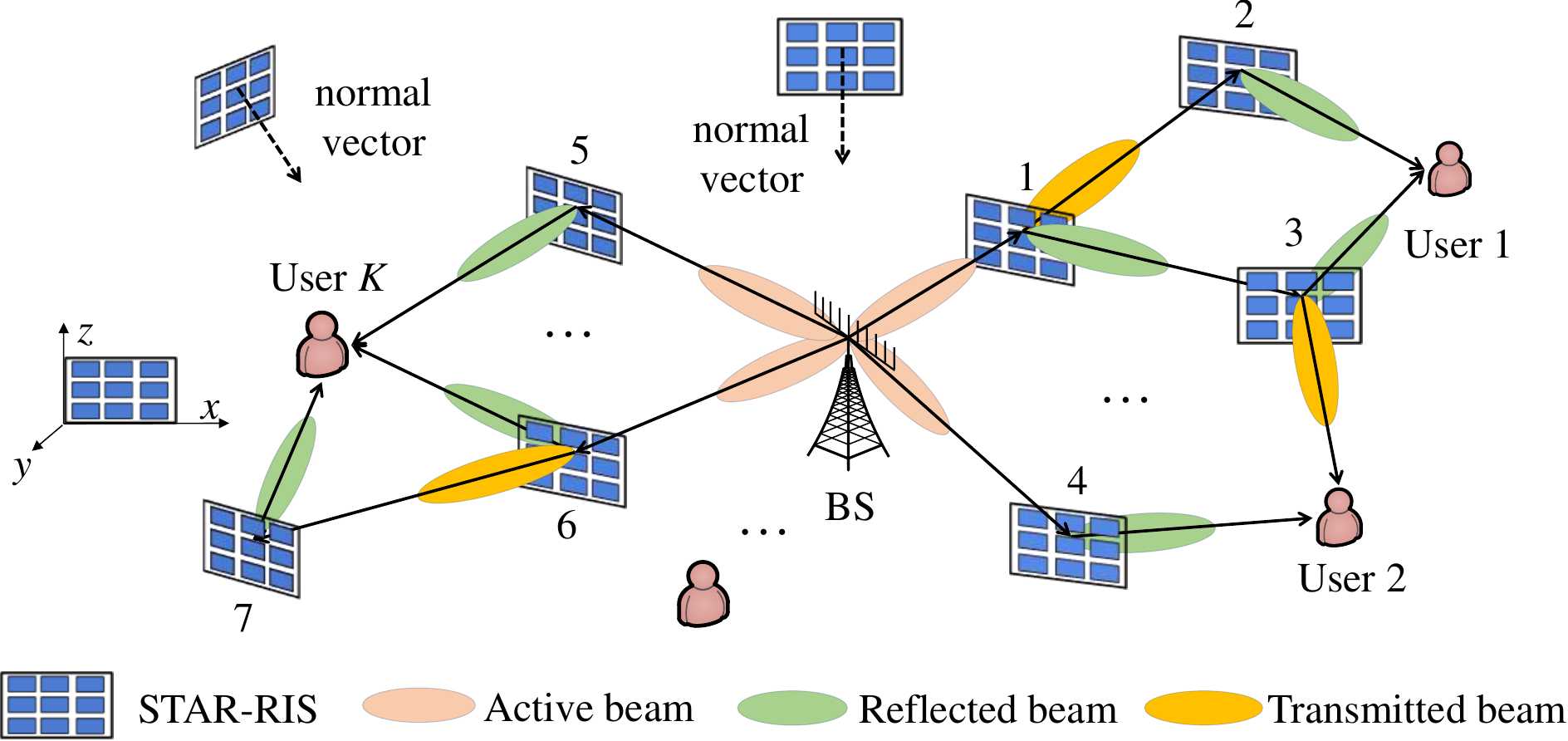}
	\caption{Multi-user multi-path beam routing with the aid of distributed STAR-RISs}\vspace{-6pt}
	\label{multicast}
\end{figure}

To fill in this gap, we investigate a multi-STAR-RIS-aided wireless communication system and its beam routing design, as shown in Fig. \ref{multicast}. In particular, we consider the multicast transmission from a multi-antenna BS to multiple single-antenna users via a set of multi-hop paths each formed by multiple STAR-RISs. The main contributions of this paper are summarized as follows.

\begin{itemize}
\item To reveal essential insights into the STAR-RIS-enabled beam routing, we consider a simplified single-user scenario. With a goal to maximize the received signal power at the user, we jointly optimize the BS’s active beamforming, the number of beam-routing paths, the BS’s power allocation over different paths, the selected STAR-RISs in each path, as well as their transmission/reflection amplitudes and phase shifts. Compared to the beam-routing optimization for conventional reflection-only RISs studied in \cite{mei2020cooperative,mei2021multi,9829192,10589431,mei2022intelligent,10643789,zhang2022multi,9900387}, the beam-routing problem for STAR-RISs is significantly more challenging due to the following reasons. First, the beam-routing paths can be intricately coupled at their shared STAR-RISs, unlike the node-disjoint path selection assumed in \cite{mei2020cooperative,mei2021multi,9829192,10589431,mei2022intelligent,10643789,zhang2022multi,9900387}. Second, even with a given path selection, optimizing the reflection and transmission amplitudes of each selected STAR-RIS introduces additional coupling challenges. To tackle these difficulties, we first develop a multi-STAR-RIS reflection/transmission signal model, based on which the optimal BS active beamforming and STAR-RIS amplitude and phase-shifts are derived in closed-form for any given path selection by employing an induction method. The derived results reveal that the maximum received signal power for a single user is equal to the sum of the maximum channel power gains over the selected paths. Furthermore, we employ a clique-based approach based on graph theory to solve the remaining multi-path selection problem efficiently via partial enumeration.
\item We extend the above STAR-RIS-enabled multi-path beam routing scheme to a more general multi-user scenario by selecting disjoint beam-routing paths for different users. Compared to the single-user case, we further optimize the power allocations for different users at the BS to maximize the minimum received signal powers among them, subject to additional constraints on the disjointness of the selected paths for different users. We derive the optimal power allocation for multiple users in closed-form given other optimization variables and apply the clique-based method to obtain a high-quality suboptimal solution for the multi-user beam-routing. Numerical results demonstrate that our proposed STAR-RIS-enabled beam routing scheme significantly outperforms the conventional beam routing schemes with reflection-only RISs for both single- and multi-user system setups by offering a more pronounced LoS path diversity gain.
\end{itemize} 

The rest of this paper is organized as follows. Section \ref{Systemmodel} presents the system model. Section \ref{singleeu} presents a simplified case with a single user and the associated beam-routing design. Section \ref{multieu} presents the more general multi-user case. Section \ref{numr} presents the numerical results to show the efficacy of the proposed schemes. Section \ref{conclusion} concludes this paper and discusses future works. 

\begin{table*}[!t]
	\label{symbol}
	\caption{{List of Main Symbols}}
	\centering
	\begin{tabular}{|c|c|c|c|}
		\hline
		$N_B$&Number of BS antennas&$K$&Number of users\\
		\hline
		$J$&Number of STAR-RISs&$M$&Number of elements per STAR-RIS\\
		\hline
		$\mathcal{K}$&Set of users&$\mathcal{J}$&Set of STAR-RISs\\
		\hline
		$d_{i,j}$&Distance between nodes $i$ and $j$&$\boldsymbol{w}_B$&Active beamforming vector of the BS\\
		\hline
		$\beta_{j}^{(R)}/\beta_{j}^{(T)}$&Reflection/transmission amplitude of STAR-RIS $j$&$\theta_{j,m}^{(R)}/\theta_{j,m}^{(T)}$&\makecell{Reflection/transmission phase-shift\\ of STAR-RIS $j$'s $m$-th element}\\
		\hline
		$\boldsymbol{\Theta}_j^{(R)}/\boldsymbol{\Theta}_j^{(T)}$&\makecell{Reflection/transmission coefficient matrix\\of STAR-RIS $j$}&$\boldsymbol{\Phi}_{j}^{(R)}/\boldsymbol{\Phi}_{j}^{(T)}$&\makecell{Reflection/transmission phase-shift matrix\\ of STAR-RIS $j$}\\
		\hline
		$G_L$&LoS graph of the system&$V_L/E_L$&Vertex/edge set of $G_L$\\
		\hline
		$e_{i,j}$&Edge from vertex $i$ to vertex $j$ in $G_L$&$\mv n_j$&Normal vector of STAR-RIS $j$\\
		\hline $\mv v_{i,j}$&Direction vector from node $i$ to node $j$&$\boldsymbol{H}_{0,j}$&Channel from the BS to STAR-RIS $j$\\
		\hline
		$\boldsymbol{S}_{i,j}$&Channel from STAR-RIS $i$ to STAR-RIS $j$&
		$\boldsymbol{g}_{j,J+k}^{H}$& Channel from STAR-RIS $j$ to user $k$\\
		\hline $\kappa(\Omega)$&LoS path gain over path $\Omega$&$h_{0,J+k}(\Omega)$&BS-user $k$ channel over path $\Omega$\\
		\hline $\hat{F}_{0,J+k}(\Omega)$&Maximum end-to-end channel power gain over path $\Omega$&$\alpha_{q}$&Total transmit power allocated to the $q$-th active beam\\
		\hline
		$\boldsymbol{w}_q$&BS beamforming vector for the $q$-th active beam&$\Omega_{q,p}$&$p$-th path associated with the $q$-th active beam\\
		\hline$a_{l}^{q,p}$&Index of the $l$-th STAR-RIS in $\Omega_{q,p}$&$L_{q,p}$&Number of selected STAR-RISs in $\Omega_{q,p}$\\
		\hline $P_q$&\makecell{Number of the paths associated with\\the $q$-th active beam}&$\mathcal{I}_j/\mathcal{U}_j$&Set of the nodes before/after STAR-RIS $j$\\
		\hline $I_j/U_j$&In/Out-degree of node $j$&$\mathcal{D}$&Set of nodes with out-degree equal to 2\\
		\hline $b_{i,j}$&Cosine of the angle between $\mv v_{i,j}$ and $\mv n_j$&\makecell{$\hat{F}_{0,J+k}(\Omega_{q,p})$}&\makecell{Maximum received signal power at user $k$ over $\Omega_{q,p}$}\\
		\hline
		\makecell{$P_{J+k}$}&\makecell{Received signal power at user $k$}&\makecell{$\tilde{P}_{J+k}$} & Maximum value of $P_{J+k}$\\
		\hline
		$\tilde{\Lambda}$&Set of all selected STAR-RISs&$\Psi_j^{(R)}/\Psi_j^{(T)}$&\makecell{Set of reflected/transmitted paths\\going through STAR-RIS $j$}\\
		\hline
		$c_l^{(j,R)}/c_l^{(j,T)}$&$l$-th reflected/transmitted path in $\Psi_j^{(R)}/\Psi_j^{(T)}$&$L_{j,R}/L_{j,T}$&\makecell{Number of reflected/transmitted paths\\going through STAR-RIS $j$}\\
		\hline
		$B_{l,\text{S}}^{(j,\mu)}$&\makecell{Amplitude product of STAR-RIS $j$'s\\subsequent nodes over path $c_l^{(j,\mu)}$}&$S$&Number of candidate paths selected for each user\\
		\hline
		$G_p$&Path graph of the considered system&$Q_k$&Number of active beams associated with user $k$\\
		\hline
		$\Omega_{k,q,p}$&\makecell{$p$-th path associated with the $q$-th active beam\\for user $k$}&$P_{k,q}$&\makecell{Number of paths associated with the $q$-th active beam\\ for user $k$}\\
		\hline
		$\boldsymbol{w}_{k,q}$&$q$-th active beam associated with user $k$&$\alpha_{k}$&Total transmit power allocated to user $k$\\
		\hline
	\end{tabular}
\end{table*}

The following notations are used in this paper. Bold symbols in capital letter and small letter denote matrices and vectors, respectively. The conjugate, transpose and conjugate transpose of a vector or matrix are denoted by $(\cdot)^*$ , $(\cdot)^T$ and $(\cdot)^H$, respectively. $\mathbb{R}_n$ ( $\mathbb{C}_n$) denotes the set of real (complex) vectors of length $n$. For a complex number $s$, $s^*$ and $\vert s\vert$ denote its conjugate and amplitude, respectively. For a diagonal matrix $\boldsymbol{A}$, $\mathrm{diag}(\mv A)$ denotes an $n\times 1$ vector that contains the $n$ diagonal elements of $\mv A$. $j$ denotes the imaginary unit, i.e., $j^2 = -1$. For two sets $\boldsymbol{A}$ and $\boldsymbol{B}$, $\boldsymbol{A}\cup\boldsymbol{B}$ denotes their union. $\emptyset$ denotes an empty set. $\angle(\mv a,\mv b)$ denotes the angle between vector $\mv a$ and vector $\mv b$. For ease of reference, the main symbols used in this paper are listed in Table \ref{symbol}.
\begingroup
\allowdisplaybreaks

\section{System Model}
\label{Systemmodel}
As shown in Fig. \ref{multicast}, we consider the multicast transmission\footnote{The proposed scheme is also applicable to the scenarios of wireless power transfer and of broadcast transmission (subjected to more stringent constraints on the path separation among the users for interference mitigation \cite{mei2021multi}).} from a BS equipped with $N_B$ antennas to $K$ single-antenna users with the aid of $J$ STAR-RISs. Due to the dense obstacles in the environment (e.g., an indoor environment with a complex topology), we assume that the BS can only transmit to each user through one or more LoS links each formed by multiple STAR-RISs, as shown in Fig. \ref{multicast}. We denote the sets of all users, STAR-RISs and the reflecting/transmitting elements per STAR-RIS as $\mathcal{K} \triangleq \{1,2,\cdots,K\}$, $\mathcal{J}\triangleq\{1,2,\cdots,J\}$ and $\mathcal{M}\triangleq\{1,2,\cdots,M\}$, respectively, and let $M_0=\sqrt{M}$ be the number of elements of each STAR-RIS per dimension. For convenience, we refer to the BS, STAR-RIS $j$ and user $k$ as nodes $0$, $j$ and $J+k$, respectively. Given a reference point at the BS and each STAR-RIS, we establish a global coordinate system for the considered system and denote the three-dimensional (3D) coordinate of the reference point of node $j$ as $\mv v_j\in\mathbb{R}^{3\times1}$. As such, the distance between any two nodes $i$ and $j$ can be expressed as $d_{i,j}=\Vert\mv v_i-\mv v_j\Vert$. Moreover, we consider that the transmitted signal can only be reflected/transmitted from one STAR-RIS (e.g., STAR-RIS $i$) to a farther STAR-RIS from the BS (e.g., STAR-RIS $j$ with $d_{0,i}<d_{0,j}$), but not vice versa, to ensure the outward signal transmission/reflection and avoid the severe multiplicative path loss\cite{pathloss}.

As shown in Fig. \ref{multicast}, we consider that all STAR-RISs are operated in the ES protocol\cite{mu2021simultaneously,9690478,10550177}, where each STAR-RIS element can reflect and transmit its incident signal at the same time. For each STAR-RIS $j$, let the amplitude and phase-shift of its $m$-th element for reflection be denoted as $\beta_{j,m}^{(R)}$ and $\theta_{j,m}^{(R)}$, respectively, with $\beta_{j,m}^{(R)}\in[0,1]$ and $\theta\in[0,2\pi], m\in\mathcal{M}$. To ease the hardware implementation, we also assume that all elements of STAR-RIS $j$ share a common amplitude for reflection, i.e., $\beta_{j,m}^{(R)}=\beta_{j}^{(R)},\, \forall m \in\mathcal{M}$. As such, the reflection coefficient matrix of STAR-RIS $j$ can be written as $\boldsymbol{\Theta}_j^{(R)}=\beta_{j}^{(R)}\boldsymbol{\Phi}_{j}^{(R)}$, where $\boldsymbol{\Phi}_{j}^{(R)}=\mathrm{diag}(\theta_{j,1}^{(R)},\theta_{j,2}^{(R)},\cdots,\theta_{j,M}^{(R)})$ denotes its reflection phase-shift matrix. Similarly, we express the transmission coefficient matrix of STAR-RIS $j$ by replacing the superscripts ``$R$'' above with ``$T$'' as $\boldsymbol{\Theta}_j^{(T)}=\beta_{j}^{(T)}\boldsymbol{\Phi}_{j}^{(T)}$, where $\boldsymbol{\Phi}_{j}^{(T)}$ denotes its transmission phase-shift matrix. We have $\beta_{j}^{(R)}+\beta_{j}^{(T)}=1, \forall j\in\mathcal{J}$.

To leverage the pronounced LoS diversity gain by densely deploying the STAR-RISs, we assume that only LoS links can be selected for the multi-hop transmission from the BS to the users via the STAR-RISs, while all NLoS links can be treated as part of environment scattering, which only have a marginal effect on the received signal power at each user in general, as numerically validated in our prior works \cite{mei2021multi,mei2021distributed}. This is because compared to the selected LoS paths, these randomly scattered paths lack a CPB gain while suffering from severe multiplicative path loss. To describe all cascade LoS paths from the BS to all users efficiently, we define a LoS graph for all nodes and their pairwise LoS links, denoted as $G_L=(V_L,E_L)$, where $V_L=\{0,1,2,\cdots,J+k\}$ and $E_L$ denote the sets of vertices and edges in $G_L$, respectively. We add an edge from vertex $i$ to vertex $j$, denoted as $e_{i,j}$, if the following two conditions are satisfied: 1) there exists a LoS path from node $i$ to node $j$, and 2) node $j$ is farther away from the BS than node $i$, i.e., $d_{0,j}>d_{0,i}$ for routing signal outwards from the BS except that node $j$ is a user. As such, each multi-hop path from the BS to user $k$ corresponds to a path from vertex 0 to vertex $J+k$ in $G_L$. For example, in Fig. \ref{multicast}, the path from the BS to user 1 going through STAR-RISs 1 and  2 corresponds to the path 0 $\rightarrow$ 1 $\rightarrow$ 2 $\rightarrow$ $J+1$.

Next, we characterize the end-to-end channel from the BS to user $k$ in any given multi-hop LoS path, denoted as $\Omega=\{a_1,a_2,\cdots,a_L\}$ with $e_{a_l,a_{l+1}}\in E_L, l=0,1,\cdots,L$, where $L\geq 1$ and $a_l$ denote the number of STAR-RISs in $\Omega$ and the index of the $l$-th STAR-RIS in this path, and we set $a_0=0$ and $a_{L+1}=J+k$. Notably, the selected STAR-RISs in $\Omega$ may use different surfaces to either reflect or transmit their respective incoming signals, depending on the locations of its previous and next nodes. For example, if nodes $a_{l-1}$ and $a_{l+1}$ are located on the same side/different sides of STAR-RIS $a_l$, its reflection/transmission surface should be used to reflect/transmit the incident signal from node $a_{l-1}$ to node $a_{l+1}$. To determine the used surface of any STAR-RIS given its previous and next nodes, we define a normal vector for each STAR-RIS $j,j\in\mathcal{J}$, denoted as $\mv n_j\in\mathbb{R}^{3\times1}$, which is perpendicular to its surface and points to the half-space where the BS is located, as shown in Fig. \ref{multicast}. Furthermore, we define the direction vector from node $i$ to node $j$ as $\mv v_{i,j}=(\mv v_j-\mv v_i)/d_{i,j}$ with $\Vert\mv v_{i,j}\Vert=1$, $i, j\in V$. Hence, the cosine of the angle between $\mv v_{i,j}$ and $\mv n_j$ can be expressed as $b_{i,j}=\mv v_{i,j}^T\mv n_j$. Then, it can be shown that nodes $a_{l-1}$ and $a_{l+1}$ are located at the same side of STAR-RIS $a_l$ if and only if $b_{a_{l-1},a_{l}}b_{a_{l+1},a_{l}}>0$, which may occur if $\angle ( \mv v_{a_{l-1},a_{l}},\mv n_{a_l} ), \angle(\mv v_{a_{l+1},a_{l}},\mv n_{a_l})\in(0,\frac{\pi}{2})$ or $\angle ( \mv v_{a_{l-1},a_{l}},\mv n_{a_l} ), \angle(\mv v_{a_{l+1},a_{l}},\mv n_{a_l})\in(\frac{\pi}{2},\pi)$. While if $b_{a_{l-1},a_{l}}b_{a_{l+1},a_{l}}<0$, they should be located at different sides of STAR-RIS $a_k$ \footnote{Note that we have assumed that any normal vector is non-orthogonal to any direction vector, i.e., $b_{a_{l-1},a_l}b_{a_l,a_{l+1}} \ne 0$, by properly deploying the STAR-RISs.}. For example, in Fig. \ref{multicast}, we have $b_{2,1}b_{3,1}<0$, indicating that STAR-RIS 2 and 3 are located at different sides of STAR-RIS 1.

Assume that the BS and each STAR-RIS are equipped with a uniform linear array (ULA) and a uniform rectangular array (URA) parallel to the $x-z$ plane, respectively, as shown in Fig. \ref{multicast}. Let $\boldsymbol{H}_{0,j}\in\mathbb{C}^{M\times N_B},j\in\mathcal{J}$ denote the channel from the BS to STAR-RIS $j$, $\boldsymbol{S}_{i,j}\in\mathbb{C}^{M\times M}$ denote that from STAR-RIS $i$ to STAR-RIS $j$, and $\boldsymbol{g}_{j,J+k}^{H}\in\mathbb{C}^{1\times M},j\in\mathcal{J}$ denote that from STAR-RIS $j$ to user $k$. Note that the above channels are applicable to both the reflection and transmission sides of each STAR-RIS. Under the far-field propagation, if $e_{0,j}\in E$, i.e., an LoS path is available from the BS to STAR-RIS $j$, the BS-STAR-RIS $j$ channel is given by $\boldsymbol{H}_{0,j}=\sqrt{\gamma}d_{0,j}^{-1}\tilde{\boldsymbol{h}}_{j,2}\tilde{\boldsymbol{h}}_{j,1}^{H}$, where $\gamma$, $\tilde{\boldsymbol{h}}_{j,1}$ and $\tilde{\boldsymbol{h}}_{j,2}$ denote the path gain at the reference distance of one meter (m), the array response from the BS to STAR-RIS $j$ and that at STAR-RIS $j$ from the BS, respectively. Similarly, if $e_{i,j}, e_{j,J+k}\in E$, the STAR-RIS $i$-STAR-RIS $j$ and STAR-RIS $j$-user $k$ channels can be expressed as $\boldsymbol{S}_{i,j}=\sqrt{\gamma}d_{i,j}^{-1}\tilde{\boldsymbol{s}}_{i,j,2}\tilde{\boldsymbol{s}}_{i,j,1}^{H}$ and $\boldsymbol{g}_{j,J+k}^{H}=\sqrt{\gamma}d_{j,J+k}^{-1}\tilde{\boldsymbol{g}}_{j,J+k}^{H}$, respectively, where $\tilde{\boldsymbol{s}}_{i,j,1}$, $\tilde{\boldsymbol{s}}_{i,j,2}$ and $\tilde{\boldsymbol{g}}_{j,J+k}$ are their corresponding array responses. Denote by $\vartheta_{0,j}$, $\vartheta^a_{i,j}/\vartheta^e_{i,j}$ and $\varphi^a_{i,j}/\varphi^e_{i,j}$ the angle-of-departure (AoD) from the BS to STAR-RIS $j$, the azimuth/elevation AoD from STAR-RIS $i$ to node $j$ (STAR-RIS or the user) and the azimuth/elevation angle-of-arrival (AoA) at STAR-RIS $j$ from node $i$ (STAR-RIS or BS), respectively. Then, assuming far-field propagation between any two nodes, we have
\begin{align}\nonumber
	&(\tilde{\boldsymbol{h}}_{j,1})_n=e^{-j2\pi(n-1)d_{A}\cos\vartheta_{0,j}/\lambda},\\\nonumber
	&(\tilde{\boldsymbol{h}}_{j,2})_m=e^{-j2\pi d_{I}(\lfloor\frac{m-1}{M_{0}}\rfloor\psi^{(1)}_{j,0}+(m-1-\lfloor\frac{m-1}{M_{0}}\rfloor M_{0})\phi^{(2)}_{j,0})/\lambda},\\\nonumber
	&(\tilde{\boldsymbol{s}}_{i,j,1})_m=e^{-j2\pi d_{I}(\lfloor\frac{m-1}{M_{0}}\rfloor\phi^{(1)}_{i,j}+(m-1-\lfloor\frac{m-1}{M_{0}}\rfloor M_{0})\phi^{(2)}_{i,j})/\lambda},\\\nonumber
	&(\tilde{\boldsymbol{s}}_{i,j,2})_m=e^{-j2\pi d_{I}(\lfloor\frac{m-1}{M_{0}}\rfloor\psi^{(1)}_{j,i}+(m-1-\lfloor\frac{m-1}{M_{0}}\rfloor M_{0})\phi^{(2)}_{j,i})/\lambda},\\\nonumber
	&(\tilde{\boldsymbol{g}}_{j,J+1})_m=e^{-j2\pi d_{I}(\lfloor\frac{m-1}{M_0}\rfloor\phi^{(1)}_{j,J+1}+(m-1-\lfloor\frac{m-1}{M_0}\rfloor M_0)\phi^{(2)}_{j,J+1})/\lambda},\vspace{-3pt}
\end{align}
where $(\mv x)_m$ denotes the $m$-th entry of a vector $\mv x$, $\phi^{(1)}_{i,j}=\sin\vartheta_{i,j}^{e}\cos\vartheta_{i,j}^{a}$, $\psi^{(1)}_{j,i}=\sin\varphi_{j,i}^{e}\cos\varphi_{j,i}^{a}$, $\phi^{(2)}_{i,j}=\cos\vartheta_{i,j}^{e}$, and $\phi^{(2)}_{j,i}=\cos\varphi_{j,i}^{e}, i, j \in V$. Moreover, $\lambda$, $d_A$, and $d_I$ denote the carrier wavelength, the BS's antenna spacing, and the STAR-RIS's element spacing, respectively.

We define $\kappa(\Omega)=(\sqrt{\gamma})^{L+1}\prod_{l=0}^Ld_{a_l,a_{l+1}}^{-1}$ and $D(\Omega) = \sum_{l=0}^{L}d_{a_{l},a_{l+1}}$ as the cascaded LoS path gain and the propagation distance from the BS to user $k$ over $\Omega$, respectively. Let $\boldsymbol{w}_B\in \mathbb{C}^{N\times1}$ denote the BS’s transmit beamforming vector with $\vert\boldsymbol{w}_B\vert^2=1$. Then, the BS-user $k$ end-to-end channel over the path $\Omega$ is given by
\begin{align}
		h_{0,J+k}(\Omega)& =\boldsymbol{g}_{a_L,J+k}^H\boldsymbol{\Phi}_{a_L}^{(\mu_L)}\left(\prod_{l=1}^{L-1}\boldsymbol{S}_{a_l,a_{l+1}}\boldsymbol{\Phi}_{a_l}^{(\mu_l)}\right)\boldsymbol{H}_{0,a_1}\boldsymbol{w}_B \nonumber\\
		&=\kappa(\Omega)e^{-j\frac{2\pi D(\Omega)}{\lambda}}\left(\prod_{l=1}^{L}A_{l}\right)(\tilde{\boldsymbol{h}}_{a_{1},1}^{H}\boldsymbol{w}_{B}),\label{pwo}
\end{align}
where 
\begin{equation}
	A_{l}=
	\begin{cases}
		\sqrt{\beta_{a_1}^{(\mu_1)}}\tilde{\boldsymbol{s}}_{a_1,a_2,1}^{H}\boldsymbol{\Phi}_{a_1}^{(\mu_1)}\tilde{\boldsymbol{h}}_{a_1,2} & \text{if}~l=1\\ \sqrt{\beta_{a_{L}}^{(\mu_L)}}\tilde{\boldsymbol{g}}_{a_{L,J+k}}^{H}\boldsymbol{\Phi}_{a_{L}}^{(\mu_L)}\tilde{\boldsymbol{s}}_{a_{L-1},a_{L},2} & \text{if}~l=L\\ \sqrt{\beta_{a_{l}}^{(\mu_l)}}\tilde{\boldsymbol{s}}_{a_{l},a_{l+1},1}^{H}\boldsymbol{\Phi}_{a_{l}}^{(\mu_l)}\tilde{\boldsymbol{s}}_{a_{l-1},a_{l},2} & \text{otherwise},
	\end{cases}\label{maxA}
\end{equation}
and
\begin{equation}
	\mu_l=
	\begin{cases}
		R, &{\text{if}}~b_{a_{l-1},a_l}b_{a_{l+1},a_l}>0\\
		T, &{\text{otherwise}}
	\end{cases}
\end{equation}
indicates the used surface of STAR-RIS $a_l$ in path $\Omega$. For example, in Fig. \ref{multicast}, the reflection surface of STAR-RIS 1 is used in the path 0 $\rightarrow$ 1 $\rightarrow$ 3 $\rightarrow$ 8 (user 1) while its transmission surface is used in the path 0 $\rightarrow$ 1 $\rightarrow$ 2 $\rightarrow$ 8 (user 1).

To maximize the BS-user $k$ channel power gain for any given transmission/reflection amplitude of the STAR-RIS over $\Omega$, i.e., $\vert h_{0,J+k}(\Omega)\vert^2$, the active BS beamforming and the phase-shifts of each STAR-RIS $a_l$ should be set as\cite{mei2020cooperative}
\begin{flalign}
	&\ \boldsymbol{w}_{B}^\star=\boldsymbol{w}_B(a_l)= e^{\frac{j2\pi D(\Omega)}{\lambda}}\tilde{\boldsymbol{h}}_{a_{1},1}/\|\tilde{\boldsymbol{h}}_{a_{1},1}\|,\label{optimalw}&
\end{flalign}
\begin{align}
	\theta_{a_{l},m}^\star&=\theta(a_{l-1},a_l,a_{l+1})\nonumber\\
	&\!=\!
	\begin{cases}\angle(\tilde{\boldsymbol{s}}_{a_1,a_2,1})_m-\angle(\tilde{\boldsymbol{h}}_{a_1,2})_m&\text{if }~l=1,\\\angle(\tilde{\boldsymbol{g}}_{a_L,J+1})_m-\angle(\tilde{\boldsymbol{s}}_{a_{L-1},a_L,2})_m&\text{if }~l=L,\\\angle(\tilde{\boldsymbol{s}}_{a_l,a_{l+1},1})_m-\angle(\tilde{\boldsymbol{s}}_{a_{l-1},a_l,2})_m&\text{otherwise},
	\end{cases}\label{optimaltheta}
\end{align}
where we define $\boldsymbol{w}_B(a_l)$ as the optimal active beamforming vector for the BS to transmit to STAR-RIS $a_1$ with $e_{0,a_1}\in E_L$ and $\theta(a_{l-1},a_l,a_{l+1})$ as the optimal phase-shift for STAR-RIS $a_l$ to reflect/transmit the beam from its previous node $a_{l-1}$ to next node $a_{l+1}$ with $e_{a_{l-1},a_l},e_{a_l,a_{l+1}}\in E_L$. Based on \eqref{optimalw} and \eqref{optimaltheta}, $\vert\tilde{\boldsymbol{h}}_{a_1,1}^H\boldsymbol{w}_B\vert$ and $\vert A_{l}\vert$ can reach their maximum values of $\sqrt{N_B}$ and $M$, respectively. Given the amplitude $\beta_{a_l}^{(\mu_l)}$, by substituting \eqref{optimalw} and \eqref{optimaltheta} into \eqref{pwo}, the maximum BS-user $k$ channel power gain over the path $\Omega$ can be obtained as 
\begin{equation}
	F_{0,J+k}(\Omega)=\vert h_{0,J+k}(\Omega)\vert^2=\kappa(\Omega)\left(\prod_{l=1}^{L}\beta_{a_l}^{(\mu_l)}\right)M^{2}N_B.\label{recevp}
\end{equation}
Note that the maximum channel power gain in \eqref{recevp} is derived under the assumption of far-field LoS channels between adjacent nodes in $\Omega$. Nevertheless, as shown in \cite{9723331}, in the case of near-field LoS channels, the far-field maximum beamforming gain can still be achieved closely by carefully optimizing the BS's active beamforming and the STAR-RISs' phase-shifts. As such, the maximum channel power gain in \eqref{recevp} can serve as an accurate approximation to that in the near-field scenario for a certain distance range. Furthermore, we denote the maximum end-to-end channel power gain of $F_{0,J+k}(\Omega)$ as
\begin{equation}
	\hat{F}_{0,J+k}(\Omega)=M^{2L}N_{B}\kappa^{2}(\Omega),
\end{equation}
which is achieved by setting $\beta_{a_l}^{(\mu_l)}=1, \mu_l\in\{R,T\}.$ Evidently, we have $F_{0,J+k}(\Omega)=\prod_{l=1}^{L}\beta_{a_l}^{(\mu_l)}\hat{F}_{0,J+k}(\Omega)$

In our previous works \cite{mei2020cooperative,mei2021multi,mei2022intelligent,10643789}, we have shown how to select multiple node-disjoint paths from the BS to the user for multi-path routing in the case with conventional reflection-only RISs. However, the STAR-RISs may be involved in multiple paths at the same time for simultaneous transmission and reflection. As such, a new multi-path beam routing scheme should be designed for STAR-RISs. Next, to reveal essential insights, we first focus on a simplified single-user case and compare the associated beam-routing design with that for conventional reflection-only RISs.

\section{Multi-path Beam Routing with Passive Beam Splitting for a Single user}
\label{singleeu}

In this section, we first focus on a simple scenario with a single user (i.e. $K=1$), as shown in Fig. \ref{esms}. In our proposed STAR-RIS-enabled multi-path beam routing, the BS splits its beamforming vector, $\boldsymbol{w}_B$, into multiple active beams via its active beam splitting, each successively transmitted and/or reflected by the selected STAR-RISs, thereby creating more split paths via passive beam splitting. Finally, the signals over all split paths are coherently combined at the user’s receiver. The details are presented below.
\vspace{-12pt}
\begin{figure}[!t]
	\centering
	\includegraphics[width=3.50in]{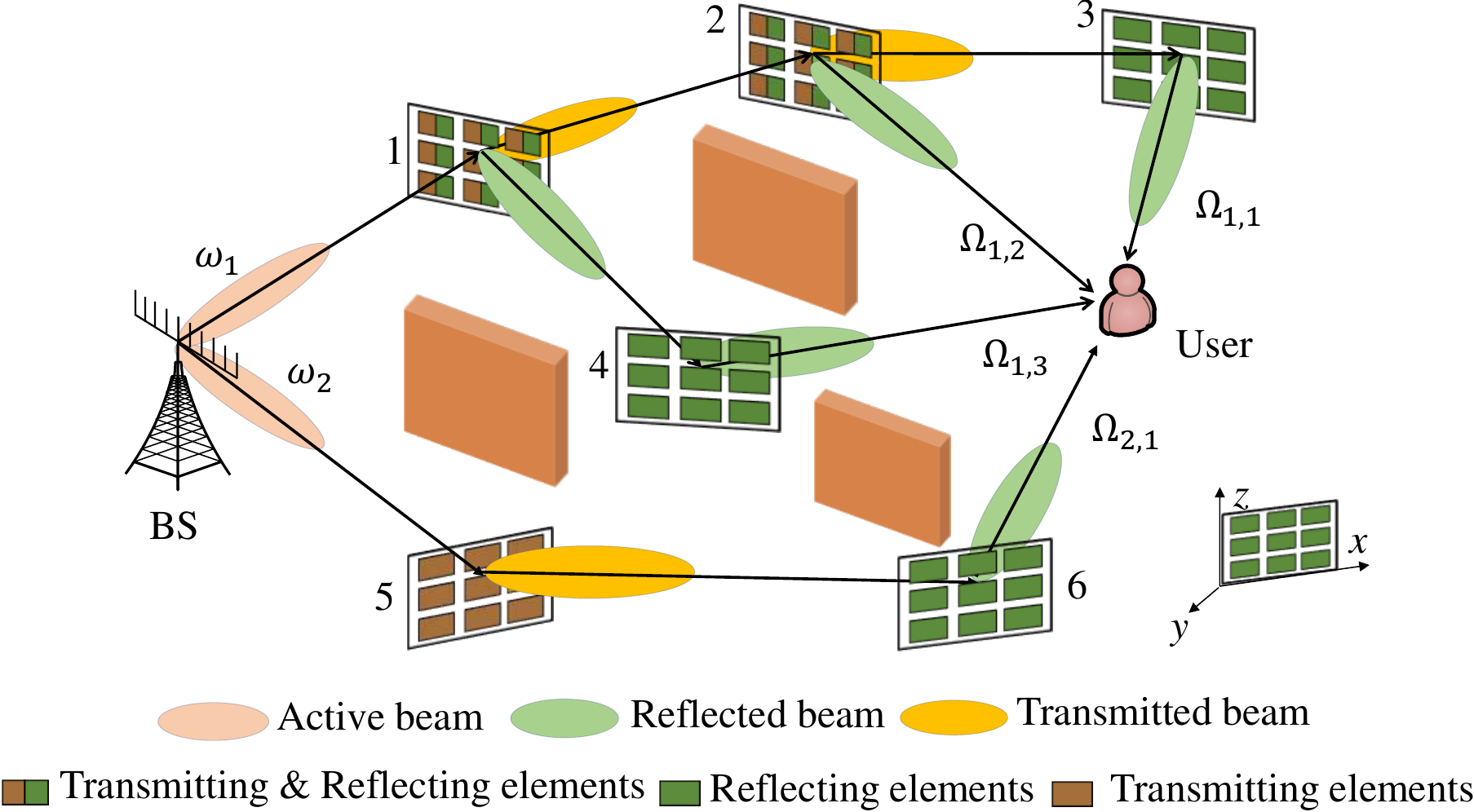}
	\caption{Multi-STAR-RIS-aided wireless communications in the single-user case.}\vspace{-12pt}
	\label{esms}
\end{figure}

\subsection{Active and Passive Beam Splitting}
\label{beamspliting}
First, we assume that the BS splits its beamforming vector into $Q(\geq1)$ beams and denote by $\boldsymbol{w}_q\in\mathbb{C}^{N\times1}$ the $q$-th split beam, $q\in\mathcal{Q}=\{1,2,\cdots,Q\}$. Let $\alpha_q$ denote the proportion of the total transmit power allocated to the $q$-th active beam with $\sum_{q=1}^{Q}\alpha_q=1$. As such, the BS's active beamforming can be expressed as $\boldsymbol{w}_B=\sum_{q=1}^{Q}\alpha_q\boldsymbol{w}_q$.

For each split active beam $\boldsymbol{w}_q$, it may be split into two beams if it is transmitted and reflected at the same time by a STAR-RIS. The resulted two passive beams may be split into more beams by other STAR-RISs, which give rise to more end-to-end LoS paths. For example, in Fig. \ref{esms}, the beam $\boldsymbol{w}_1$ is split at STAR-RIS 1 into two beams, one of which is further split at STAR-RIS 2, thus creating three LoS paths from the BS to the user, i.e., $\Omega_{1,1}$, $\Omega_{1,2}$ and $\Omega_{1,3}$. As such, the beam $\boldsymbol{w}_1$ is associated with these three paths. Assume that each split active beam $\boldsymbol{w}_q$ is associated with $P_q$ paths and denote the $p$-th path as $\Omega_{q,p}=\{a_1^{(q,p)},a_2^{(q,p)},\cdots,a_{L_{q,p}}^{(q,p)}\}$, $p\in\{1,2,\cdots,P_q\}$, where $a_l^{(q,p)}$ and $L_{q,p}$ denote the index of the $l$-th used STAR-RIS in $\Omega_{q,p}$ and the total number of selected STAR-RISs in this path, respectively. 

Let the set of all selected STAR-RISs be denoted as $\tilde{\Lambda}=\cup_{q,p}\Omega_{q,p}$ and define $\Lambda=\tilde{\Lambda}\cup\{0,J+1\}$. For each STAR-RIS $j, j\in\tilde{\Lambda}$, we denote the set of its previous nodes and next nodes as $\mathcal{I}_j=\{i\mid e_{i,j}\in E_L, i\in\Lambda\}$ and $\mathcal{U}_j=\{i\mid e_{j,i}\in E_L, i\in\Lambda\}$, respectively. As such, we can express the number of previous and next nodes of STAR-RIS $j$, $j\in\tilde{\Lambda}$ (a.k.a. the in-degree and out-degree of node $j$) as $I_j=\vert\mathcal{I}_j\vert$ and $U_j=\vert\mathcal{U}_j\vert$, respectively. To simplify the optimization, we assume that each selected STAR-RIS only reflect or/and transmit an incident signal from one previous node to one or two next nodes located at its different sides, i.e.,
\begin{equation}
	I_j=1, U_j\in\{1,2\},\quad \forall j\in \tilde{\Lambda}.\label{inoutd}
\end{equation}
It is worth noting that only when STAR-RIS $j$ reflects and transmits its incident signal at the same time, we have $U_j=2$. Let $\mathcal{D}=\{j\mid U_j=2,j\in\tilde{\Lambda}\}$ denote the set of selected STAR-RISs with their out-degree equal to 2. For example, in Fig. \ref{esms}, we have $\mathcal{D}=\{1,2\}$. For each STAR-RIS $j,j\in\mathcal{D}$, as its two next nodes should be located at its different sides, it must hold that
\begin{equation}
	b_{i,j}b_{k,j}<0,\quad\forall i,k\in\mathcal{U}_{j}, i\neq k. j\in\mathcal{D}.\label{D}
\end{equation} 
\vspace{-9pt}

\subsection{Optimal STAR-RIS Phase Shifts and BS Power Allocation}
\label{optimalwandt}
Next, we show that given the paths $\Omega_{q,p}$, $q \in {\cal P}_q$, $q \in {\cal Q}$, we can obtain the optimal STAR-RIS phase shifts for transmission/reflection (i.e., $\boldsymbol{\Phi}_j^{(T)}$ and $\boldsymbol{\Phi}_j^{(R)}$, $j \in \Lambda$) and the BS power allocation (i.e., $\alpha_q, q \in \cal Q$) in closed form. Specifically, given \eqref{inoutd} and \eqref{D}, it is noted that the transmission and reflection phase shifts of each STAR-RIS can be determined only based on its channels with one previous node and one next node, respectively, similarly to \eqref{optimaltheta}. For example, in Fig. \ref{esms}, the transmission phase shifts of STAR-RIS 2 only depend on STAR-RISs 1 and 3, while its reflection phase shifts only depend on STAR-RIS 1 and the user. As such, the optimal phase shifts of the STAR-RISs in $\Omega_{q,p}$ can be obtained based on \eqref{optimaltheta} by replacing the nodes in $\Omega$ with those in $\Omega_{q,p}$, and the resulting channel power gain can be obtained as $\alpha_qF(\Omega_{q,p})$ similar to \eqref{recevp}. By setting the reflection/transmission amplitude of all STAR-RISs in $\Omega_{q,p}$ to unity, its maximum channel power gain is given by $\alpha_q\hat{F}(\Omega_{q,p})$.

Notably, the signals over the paths $\Omega_{q,p},\, q\in\mathcal{Q},p\in\mathcal{P}_q$ may not coherently add at the user’s receiver because of their different time delays causing random channel phases. To maximize the received power, a phase rotation $e^{-\angle(h_{0,J+1}(\Omega_{q,p}))}$ should be appended to any STAR-RIS in the path $\Omega_{q,p}$ to ensure that signals over different paths are coherently combined. The value of rotated phases can be computed by the BS based on distributed beam training\cite{mei2021distributed}. As such, the received power at the user can be expressed as
\begin{equation}
	P_{J+1}=\left(\sum_{q=1}^{Q}\sum_{p=1}^{P_{q}}\sqrt{\alpha_{q}F_{0,J+1}(\Omega_{q,p})}\right)^{2}\leq\sum_{q=1}^{Q}\Gamma_q^{2}\triangleq\tilde{P}_{J+1},\label{eup}
\end{equation}
where $\Gamma_q=\sum_{p=1}^{P_q}\sqrt{F_{0,J+1}(\Omega_{q,p})}$ and the inequality is according to the Cauchy-Schwarz inequality. Note that to make the equality in \eqref{eup} hold, we should set
\begin{equation}
	\alpha_{q}=\frac{\Gamma_{q}^2}{\sum_{q=1}^{Q}\Gamma_{q}^2},\  q\in\mathcal{Q}.
\end{equation} 
\vspace{-12pt}

\subsection{Optimal STAR-RIS Amplitudes}\label{es}
Next, we derive the optimal reflection and transmission amplitude of each STAR-RISs in $\Lambda$. We start from the simple example shown in Fig. \ref{esms}(a), where the BS splits its beamforming vector into $Q=2$ active beams (i.e. $\boldsymbol{w}_1$ and $\boldsymbol{w}_2$) transmitted to the user via the STAR-RISs deployed between them. The active beam $\boldsymbol{w}_1$ is further split by STAR-RIS 1 and STAR-RIS 2, resulting in three split paths denoted as $\Omega_{1,1}=\{1,2,3\}$, $\Omega_{1,2}=\{1,2\}$ and $\Omega_{1,3}=\{1,4\}$. Thus, we have $\Gamma_1=\sqrt{F_{0,J+1}(\Omega_{1,1})}+\sqrt{F_{0,J+1}(\Omega_{1,2})}+\sqrt{F_{0,J+1}(\Omega_{1,3})}$ with $F_{0,J+1}(\Omega_{1,1}) = \beta_{1}^{(T)}\beta_{2}^{(T)}\beta_{3}^{(R)}\hat{F}_{0,J+1}(\Omega_{1,1})$, $F_{0,J+1}(\Omega_{1,2}) = \beta_{1}^{(T)}\beta_{2}^{(R)}\hat{F}_{0,J+1}(\Omega_{1,2})$ and $F_{0,J+1}(\Omega_{1,3}) = \beta_{1}^{(R)}\beta_{4}^{(R)}\hat{F}_{0,J+1}(\Omega_{1,3})$. Similarly, for the active beam $\boldsymbol{w}_2$, we have $\Omega_{2,1}=\{5,6\}$ and $\Gamma_2=\sqrt{F_{0,J+1}(\Omega_{2,1})}$ where $F_{0,J+1}(\Omega_{2,1}) =\beta_{5}^{(T)}\beta_{6}^{(R)}\hat{F}_{0,J+1}(\Omega_{2,1})$. Note that for the STAR-RISs that are only involved in a single path, their reflect/transmit amplitude should be set to unity to maximize the received power at the user, as either their reflection or transmission surfaces are used. For example, in Fig. \ref{esms}, we set $\beta_{3}^{(R)}$, $\beta_{4}^{(R)}$, $\beta_{5}^{(T)}$ and $\beta_{6}^{(R)}$ to unity. By substituting the above into \eqref{eup}, the received power at the user can be expressed as
	\begin{align}
		P_{J+1} =&\left(\sqrt{F_{0,J+1}(\Omega_{1,1})}+\sqrt{F_{0,J+1}(\Omega_{1,2})}\right.\nonumber\\&\left.\qquad\qquad+\sqrt{F_{0,J+1}(\Omega_{1,3})}\right)^{2}+F_{0,J+1}(\Omega_{2,1})\nonumber \\
		=&\left(\sqrt{\beta_{1}^{(R)}\hat{F}_{0,J+1}(\Omega_{1,3})}+\sqrt{\beta_{1}^{(T)}\beta_{2}^{(T)}\hat{F}_{0,J+1}(\Omega_{1,1})}\right.\nonumber\\&\left.+\sqrt{\beta_{1}^{(T)}\beta_{2}^{(R)}\hat{F}_{0,J+1}(\Omega_{1,2})}\right)^{2}+\hat{F}_{0,J+1}(\Omega_{2,1})\nonumber\\
		\leq &\hat{F}_{0,J+1}(\Omega_{2,1})+\left(\beta_{1}^{(T)}\beta_{2}^{(T)}+\beta_{1}^{(T)}\beta_{2}^{(R)}+\beta_{1}^{(R)}\right)\nonumber\\&\times\left(\hat{F}_{0,J+1}(\Omega_{1,1})+\hat{F}_{0,J+1}(\Omega_{1,2})+\hat{F}_{0,J+1}(\Omega_{1,3})\right)\nonumber
		\\=&\hat{F}_{0,J+1}(\Omega_{1,1})+\hat{F}_{0,J+1}(\Omega_{1,2})+\hat{F}_{0,J+1}(\Omega_{1,3})\nonumber\\&+\hat{F}_{0,J+1}(\Omega_{2,1})\triangleq\tilde{P}_{J+1},\label{maxpexample}
	\end{align}
where the inequality is due to the Cauchy-Schwarz inequality, and the last equality is due to the fact that $\beta_1^{(T)}+\beta_1^{(R)}=1$ and $\beta_2^{(T)}+\beta_2^{(R)}=1$. Moreover, it can be shown that by setting the amplitudes of each STAR-RIS $j$, $j\in\mathcal{D}$ as follows, the equality in \eqref{maxpexample} can always hold.
\begin{align}
	\beta_{1}^{(R)}=&\frac{{\hat F}_{0,J+1}(\Omega_{1,3})}{\sum_{p=1}^{3}\hat F_{0,J+1}(\Omega_{1,p})},\nonumber\\
	\beta_{1}^{(T)}=&\frac{{\hat F}_{0,J+1}(\Omega_{1,1})+{\hat F}_{0,J+1}(\Omega_{1,2})}{\sum_{p=1}^{3}\hat F_{0,J+1}(\Omega_{1,p})},\nonumber\\
	\beta_{2}^{(R)}=&\frac{{\hat F}_{0,J+1}(\Omega_{1,2})}{{\hat F}_{0,J+1}(\Omega_{1,1})+{\hat F}_{0,J+1}(\Omega_{1,2})},\nonumber\\
	\beta_{2}^{(T)}=&\frac{{\hat F}_{0,J+1}(\Omega_{1,1})}{{\hat F}_{0,J+1}(\Omega_{1,1})+{\hat F}_{0,J+1}(\Omega_{1,2})}.\label{optimalamplitudeex}
\end{align}

The result in \eqref{maxpexample} indicates that the maximum received signal power with the optimal amplitude of the selected STAR-RISs is equal to the sum of the channel power gains over the paths $\Omega_{1,1}$, $\Omega_{1,1}$, $\Omega_{1,3}$, and $\Omega_{2,1}$. Next, we extend this observation to a general setup. To this end, we first define any end-to-end path $\Omega_{q,p}$ as a transmitted/reflected path for STAR-RIS $j, j \in \Omega_{q,p}$, if this path goes through its transmission/reflection surface. For example, in Fig. \ref{esms}, $\Omega_{1,1}$ and $\Omega_{1,2}$ are the transmitted paths of STAR-RIS 1 while $\Omega_{1,3}$ is its reflected path. Accordingly, we can denote the set of reflected paths and transmitted paths for each STAR-RIS $j$, $j\in\Lambda$ as $\Psi_j^{(R)}=\{c_1^{(j,R)},c_2^{(j,R)},\cdots c_{L_{j,R}}^{(j,R)}\}$ and $\Psi_j^{(T)}=\{c_1^{(j,T)},c_2^{(j,T)},\cdots c_{L_{j,T}}^{(j,T)}\}$, respectively, where $c_l^{(j,\mu)}$ and $L_{j,\mu}$, $\mu\in\{R,T\}$ denote the $l$-th reflected/transmitted path and the number of reflected/transmitted paths going through STAR-RIS $j$, respectively. Note that if only the reflect or transmit surface of STAR-RIS $j$ is used, the set $\Psi_j^{(T)}$ or $\Psi_j^{(R)}$ should be set to $\emptyset$, respectively. To proceed, we propose the following proposition.

\begin{proposition}
	\label{Prop1}
Given all paths $\Omega_{q,p}$'s from the BS to the user and subject to \eqref{inoutd} and \eqref{D}, the maximum received signal power is equal to the sum of channel power gains over all selected paths, which is given by
	\begin{equation}
		\tilde{P}_{J+1}=\sum_{q=1}^{Q}\sum_{p=1}^{P_q}\hat{F}_{0,J+1}(\Omega_{q,p}),\label{proposition1pw}
	\end{equation}
	and the optimal amplitude of each selected STAR-RIS $j$, $j\in\tilde{\Lambda}$, should be set as
	\begin{equation}
		\beta_{j}^{(\mu)}=\frac{\sum_{l=1}^{L_{j,\mu}}\hat F(c^{(j,\mu)}_l)}{\sum_{l=1}^{L_{j,T}}\hat F(c^{(j,T)}_l)+\sum_{l=1}^{L_{j,R}}\hat F(c^{(j,R)}_l)},\quad \mu\in\{R,T\}.\label{proposeoptimalbeta}
	\end{equation}
\end{proposition}

\begin{IEEEproof}
	Proposition \ref{Prop1} can be proved via induction, and the details are provided in Appendix \ref{Proof1}.
\end{IEEEproof}

\textbf{Proposition \ref{Prop1}} indicates that the maximum received signal power is equal to the sum of the maximum channel power gains of the selected paths, as if they were node disjoint. This suggests that the enhanced LoS path diversity gain and the coherent beam combining at the user can effectively compensate for the loss of the maximum CPB gain over each path resulted from the passive beam splitting. The result in \eqref{proposition1pw} also greatly simplifies the remaining path selection optimization, as detailed.

\begin{remark}
	Note that the Cauchy-Schwartz inequality has also been applied in our previous works (e.g., \cite{mei2022intelligent,10643789}) for multi-path beam routing. However, it is only used at the BS for an {\it active} beam splitting design. In contrast, in this paper, it is further used for {\it passive} beam splitting at multiple STAR-RISs coupled over different paths. Notably, the passive beam splitting design for one STAR-RIS can affect those for others, rendering the overall beam splitting optimization significantly more challenging. 
\end{remark}\vspace{-6pt}

\subsection{Problem Formulation}
In this paper, we aim to jointly optimize the number of the active beams ($Q$) and their associated paths ($P_q$, $q\in\cal Q$), as well as the selected STAR-RISs in each of these paths ($\Omega_{q,p}$'s) to maximize the received signal power in \eqref{proposition1pw}. The associated optimization problem is formulated as 
\begin{equation}
	\begin{aligned}\nonumber
		\left( \mathrm{P1}\right) &\operatorname*{max}_{P_q,Q,\{\Omega_{q,p}\}_{q\in\mathcal{Q},p\in\mathcal{P}_q}}\sum_{q=1}^{Q}\sum_{p=1}^{P_q}\hat F_{0,J+1}(\Omega_{q,p})\\
		&s.t.\quad e(a_l^{(q,p)},a_{l+1}^{(q,p)})\in E_L,\:\\&\:\:\quad\quad\forall l=0,1,\cdots,L_{q,p},\:p\in\mathcal{P}_q,\:q\in\mathcal{Q},\\&\:\:\quad\quad\eqref{inoutd},\:\eqref{D}.
	\end{aligned}
\end{equation}

It is noted that if we constrain $I_j=U_j=1$ and remove \eqref{D} in $\mathrm{(P1)}$, all selected beam-routing paths will become node-disjoint, under which each selected STAR-RIS either reflects or transmits the incident signal, i.e., it works under the mode selection (MS) protocol\cite{mu2021simultaneously,9690478,10550177}. As such, the optimal value of $\mathrm{(P1)}$ should be no smaller than that achieved by the MS protocol. However, $\mathrm{(P1)}$ is a more challenging combinatorial optimization problem due to the intricate coupling of the paths at their shared STAR-RISs in $\cal D$ (instead of being node-disjoint as in \cite{mei2022intelligent}). Next, we propose an efficient graph-based solution to solve it.

\subsection{Proposed Solution to (P1)}
\label{ssolution}
Specifically, we adopt a clique-based approach in graph theory to solve (P1) via partial enumeration. First, among all paths from vertex $0$ to vertex $J+1$ in $G_L$, we select $S$ candidate paths that achieve the $S$ largest maximum end-to-end channel power gains, i.e., ${\hat F}(\Omega)$. To this end, we can define a weight $W_{i,j}=\ln\frac{d_{i,j}}{M}$ for each $e_{i,j}$ in $E_L$. Then, it can be shown that finding the $S$ paths with the largest maximum end-to-end channel power gain is equivalent to finding the $S$ paths with the smallest sum of edge weights in $G_L$ \cite{mei2020cooperative}. This is a classical graph-theoretical problem that can be optimally solved by invoking Yen's algorithm \cite{west2001introduction}, which incurs the complexity in the order of $\mathcal O (SJ(\lvert E\vert+J\log(J)))$ \cite{west2001introduction}. Denote by $\cal S$ the set of all candidate paths. From the paths in $\cal S$, we aim to find the best set of paths that is feasible to (P1) and achieves the maximum objective value.

To achieve this purpose, we can construct a path graph $G_p=(V_p, E_p)$, where each vertex in $V_p$ corresponds to one candidate paths in $\cal S$. To ensure that any selected paths satisfy the constraints in (P1), we add an edge to any two vertices in $V_p$ if and only if the nodes of their corresponding paths in $G_p$ satisfy the constraints in (P1). To proceed, we present the definitions of clique and maximal clique in the graph theory.

\begin{definition}
	\label{definicli}
	A clique refers to a subset of vertices of an undirected graph such that every two distinct vertices in the clique are adjacent.
\end{definition}

\begin{definition}
	\label{definimaxc}
	A maximal clique refers to a clique that cannot be extended by including one more adjacent vertex, that is, a clique which does not exist exclusively within the vertex set of a larger clique.
\end{definition}

Based on \textbf{Definition \ref{definicli}}, a clique with $U$ vertices corresponds to a feasible solution to (P1) with $U$ paths. Hence, we can enumerate all cliques in $G_p$ to obtain all paths sets and select the best one. However, based on \textbf{Definition \ref{definimaxc}}, there is no need to enumerate all possible cliques and we only need to enumerate the maximal cliques in $G_p$. This is because for any clique in $G_p$ (e.g., $C$), if it is not a maximal clique, there must exist another clique $C_0$ with $C\subset C_0$, which means that the corresponding paths in $C$ should also be a subset of those in $C_0$. Hence, the latter must yield a better performance than the former because more paths are constructed to serve the user. We apply the Bron-Kerbosch algorithm \cite{west2001introduction} to enumerate the maximal cliques in $G_p$, with the required worst-case computational complexity in the order of $3^{S/3}$. Note that if the number of the candidate paths $S$ is sufficiently large, the optimal path solution to $\mathrm{(P1)}$ may be included in $\cal S$. In this case, the proposed approach can achieve globally optimal performance, at the cost of higher time complexity. Fortunately, as will be shown in Section \ref{Simulations} via simulation, it generally only requires a small $S$ to achieve the optimal performance. Finally, we compare the objective value of $\mathrm{(P1)}$ achieved by each maximal clique in $G_p$ and select the corresponding paths of the best one as the output.\vspace{-9pt}

\subsection{Implementation Issues}
\subsubsection{Deployment and Cost-Performance Trade-off}
For the considered multi-STAR-RIS system, there generally exists a fundamental cost-performance trade-off. Deploying more STAR-RISs can create more split signal paths to enhance coverage, but at the expense of increased control and coordination complexity. In certain scenarios, deploying conventional lower-cost reflection-only RISs may be sufficient to achieve global LoS coverage. However, STAR-RISs help further enhance the user's received signal power due to their passive beam splitting capability, as analytically demonstrated in Proposition \ref{Prop1}. To optimally balance cost and communication performance (both between STAR-RISs themselves and between conventional RISs and STAR-RISs), a hybrid deployment of multiple RIS and STAR-RISs can be optimized using a graph-based algorithm proposed in our previous works \cite{10086045,10439018}. In this paper, we primarily focus on signal-level optimization for a given deployment of multiple STAR-RISs to characterize the fundamental performance limits, which in turn can guide the design of optimal deployment strategies.

\subsubsection{Channel State Information (CSI) Acquisition}
As only LoS-dominant links between any two nodes (if any) are utilized in the beam-routing design, we can acquire all required CSI by implementing a similar distributed beam training scheme proposed in our prior work \cite{mei2021distributed}, where the controller of each STAR-RIS is involved. Specifically, for any triple of nodes ($i$,$j$,$k$) with $e_{i,j}, e_{j,k} \in E_L$, we can let node $i$ (BS or the controller of STAR-RIS $i$) and node $k$ (the controller of STAR-RIS $i$ or the user) transmit and receive a beacon signal, respectively. In the meanwhile, STAR-RIS $j$ performs beam scanning over the entire angular domain by adjusting its reflection/transmission coefficients. By collecting and comparing the received signal power at node $k$, we can estimate the angle information on the cascaded LoS link for this triple of nodes ($i$,$j$,$k$). Note that if node $k$ is a STAR-RIS, the angle information can be obtained and fed back to the BS \textit{offline}. In real-time stage, only final-hop STAR-RISs need to perform beam scanning with their respectively adjacent users for angle estimation. By collecting the estimated angle information at each STAR-RIS in both offline and real-time stages, the BS can determine the optimal amplitude at each selected STAR-RIS based on \eqref{proposeoptimalbeta} and their multi-hop LoS paths based on the algorithm presented in Section \ref{ssolution}.

\section{Proposed Scheme for Multiple Users}\label{multieu}
In this section, we extend the proposed scheme in Section \ref{singleeu} to the multicast scenario with multiple users.

\subsection{Optimal Phase Shifts and Amplitude for Multiple STAR-RISs}
First, we assume that the BS splits its active beamforming vector into $Q\:(\geq K)$ beams, and each user $k,k\in\mathcal{K}$ is associated with $Q_k$ active beams each going through $P_{k,q}$ paths, $q\in\{1,\cdots,Q_k\}$, with $Q=\sum_{k=1}^{K}Q_k$. Denote the $p$-th path of the $q$-th active beam for user $k$ as $\Omega_{k,q,p}=\{a_1^{(k,q,p)},a_2^{(k,q,p)},\cdots,a_{L_{k,q,p}}^{(k,q,p)}\}$, $p\in{\cal P}_{k,q} \triangleq\{1,2,\cdots,P_{k,q}\}$ where $a_l^{(k,q,p)}$ and $L_{k,q,p}$ denote the index of the $l$-th STAR-RIS in $\Omega_{k,q,p}$ and the total number of selected STAR-RISs in this path, respectively. To simplify the optimization and facilitate the use of the proposed scheme in Section \ref{singleeu}, we assume that the selected paths for different users are node-disjoint, i.e., $\Omega_{k,q,p}\cap\Omega_{k',q',p'} = \emptyset$ if $k\ne k'$,\footnote{Although allowing node-joint paths could potentially lead to more efficient STAR-RIS reuse, the optimization problem becomes significantly more complex to tackle. In this case, both users and paths can be coupled at a STAR-RIS, which requires more sophisticated optimization technique to handle it.} while the paths selected for each user can be intersected subjected to \eqref{inoutd} and \eqref{D}. Given the above constraints and the paths $\Omega_{k,q,p}$, $k\in\mathcal{K}$, $q\in\mathcal{Q}_k$, $p\in\mathcal{P}_{k,q}$, it follows that the reflection/transmission amplitude and phase-shifts designs of STAR-RISs associated with different users can be decoupled. As such, we can obtain the optimal reflection/transmission phase-shifts of each STAR-RIS in a path $\Omega_{k,q,p}$ similarly to \eqref{optimaltheta} by replacing the nodes in $\Omega$ with those in $\Omega_{k,q,p}$.

Specifically, we define $\boldsymbol{w}_{k,q}\in\mathbb{C}^{N\times1}$ as the $q$-th active split beam transmitted to user $k$. Let $\alpha_k$ and $\alpha_{k,q}$ denote the proportion of the total transmit power allocated to the $k$-th user and the $q$-th active beam of user $k$, respectively, with $\alpha_k=\sum_{q=1}^{Q_k}\alpha_{k,q}$ and $\sum_{k=1}^{K}\alpha_k=1$. As such, the BS's active beamforming vector can be expressed as $\boldsymbol{w}_B=\sum_{k=1}^{K}\sum_{q=1}^{Q_k}\alpha_{k,q}\boldsymbol{w}_{k,q}$. Since all users share no common STAR-RISs and paths, Proposition \ref{Prop1} should hold for each user for any selected paths. As a result, the maximum received signal power at user $k$ for any given path selection can be expressed as
\begin{equation}
	\tilde{P}_{J+k}=\alpha_k\sum_{q=1}^{Q_k}\sum_{p=1}^{P_{k,q}}\hat{F}_{0,J+k}(\Omega_{k,q,p}), k \in {\cal K}.\label{mproposition1pw}
\end{equation} 
\vspace{-12pt}
\subsection{Problem Formulation}
In the multicast scenario, we aim to maximize the minimum received signal power among all users by jointly optimizing the selected STAR-RISs, i.e., $\Omega_{k,q,p}$'s, the number of active beams for each user $(Q_k)$ and their associated paths $(P_{k,q})$, as well as the multi-user power allocation $(\alpha_k)$. The corresponding optimization problem is formulated as
\begin{align}
	\left( \mathrm{P2}\right) &\operatorname*{max}_{\{P_{k,q}\},\{Q_k,\alpha_k\},\{\Omega_{k,q,p}\}}\operatorname*{min}_{k\in\mathcal{K}}\sum_{q=1}^{Q_k}\sum_{p=1}^{P_{k,q}}\alpha_{k}\hat{F}_{0,J+k}(\Omega_{k,q,p})\nonumber\\&s.t.\quad e(a_l^{(k,q,p)},a_{l+1}^{(k,q,p)})\in E_L,\:\forall l=0,1,\cdots,L_{k,q,p},\nonumber\\&\:\:\quad\quad p\in\mathcal{P}_{k,q},\:q\in\mathcal{Q}_k,\:k\in\mathcal{K}\nonumber\\&\:\:\quad\quad\eqref{inoutd},\:\eqref{D},\:\sum_{k=1}^{K}\alpha_k=1,\\&\:\:\quad\quad \Omega_{k,q,p}\cap\Omega_{k',q',p'} = \emptyset,\: \forall k\ne k', q\in\mathcal{Q}_k, q'\in\mathcal{Q}_{k'},\nonumber\\&\:\:\quad\quad p\in\mathcal{P}_{k,q}, p'\in\mathcal{P}_{k',q'}\nonumber.
\end{align}

Compared to (P1), (P2) is more challenging to be optimally solved due to the presence of more users and the additional constraints on the disjointness of their selected paths and power allocations. On the other hand, compared to the multi-user beam routing problems studied in our previous works\cite{mei2022intelligent,mei2021multi,10643789} for conventional reflection-only RISs, (P2) is also more complicated due to the coupled paths at their shared STAR-RISs. In the following, we propose a clique-based solution to (P2) via partial enumeration as well.
\vspace{-9pt}

\subsection{Proposed Solution to (P2)}
First, we determine the optimal power allocation solution to (P2), i.e., $\alpha_k, k \in {\cal K}$, for any given path selection. It can be shown that at the optimality of (P2), it must hold that the received signal powers of all users, i.e. $\sum_{q=1}^{Q_k}\sum_{p=1}^{P_{k,q}}\alpha_{k}\hat{F}_{0,J+k}(\Omega_{k,q,p})$, are identical. Based on this fact, we can obtain
\begin{equation}
	\alpha_{k}=\frac{\sum_{q=1}^{Q_k}\sum_{p=1}^{P_{k,q}}\hat{F}^{-1}_{0,J+k}(\Omega_{k,q,p})}{\sum_{k=1}^{K}\sum_{q=1}^{Q_k}\sum_{p=1}^{P_{k,q}}\hat{F}^{-1}_{0,J+k}(\Omega_{k,q,p})},\label{pwallo}
\end{equation} 
which results in
\begin{equation}
	\label{optimalmp}
	\tilde{P}_{J+k}=\frac{1}{\sum_{k=1}^{K}\sum_{q=1}^{Q_k}\sum_{p=1}^{P_{k,q}}\hat{F}^{-1}_{0,J+k}(\Omega_{k,q,p})}.
\end{equation}

Based on \eqref{optimalmp}, (P2) can be equivalently recast as
\begin{align}
	(\mathrm{P3}) &\operatorname*{min}_{\{P_{k,q}\},\{Q_k\},\{\Omega_{k,q,p}\}}\sum_{k=1}^{K}\sum_{q=1}^{Q_k}\sum_{p=1}^{P_{k,q}}\hat F^{-1}_{0,J+k}(\Omega_{k,q,p})\nonumber\\&s.t.\quad e(a_l^{(k,q,p)},a_{l+1}^{(k,q,p)})\in E,\:\forall l=0,1,\cdots,L_{k,q,p},\nonumber\\&\:\:\quad\quad p\in\mathcal{P}_{k,q},\:q\in\mathcal{Q}_k,\:k\in\mathcal{K}\nonumber\\&\:\:\quad\quad\eqref{inoutd},\:\eqref{D},\\&\:\:\quad\quad \Omega_{k,q,p}\cap\Omega_{k',q',p'} = \emptyset,\: \forall k\ne k', q\in\mathcal{Q}_k, q'\in\mathcal{Q}_{k'},\nonumber\\&\:\:\quad\quad p\in\mathcal{P}_{k,q}, p'\in\mathcal{P}_{k',q'}.\nonumber
\end{align}

To solve (P3), we adopt a similar clique-based algorithm to Section \ref{ssolution}. Specifically, for each user $k$, $k\in\mathcal{K}$, we first select $S$ candidate paths from vertex 0 to vertex $J+k$ with the $S$ largest maximum end-to-end power gains by invoking Yen's algorithm\cite{west2001introduction}, which incurs the complexity in the order of $\mathcal{O}(KSJ(\lvert E\vert+J\log(J)))$. Denote by $\mathcal{S}_k$ and $\mathcal{S}$ the set of candidate paths for user $k$ and the set of all candidate paths with $\mathcal{S}=\cup_{k=1}^K\mathcal{S}_k$.
 
\begin{figure}[!t]
	\centering
	\subfigure[3D plot.]{\includegraphics[width=0.35\textwidth]{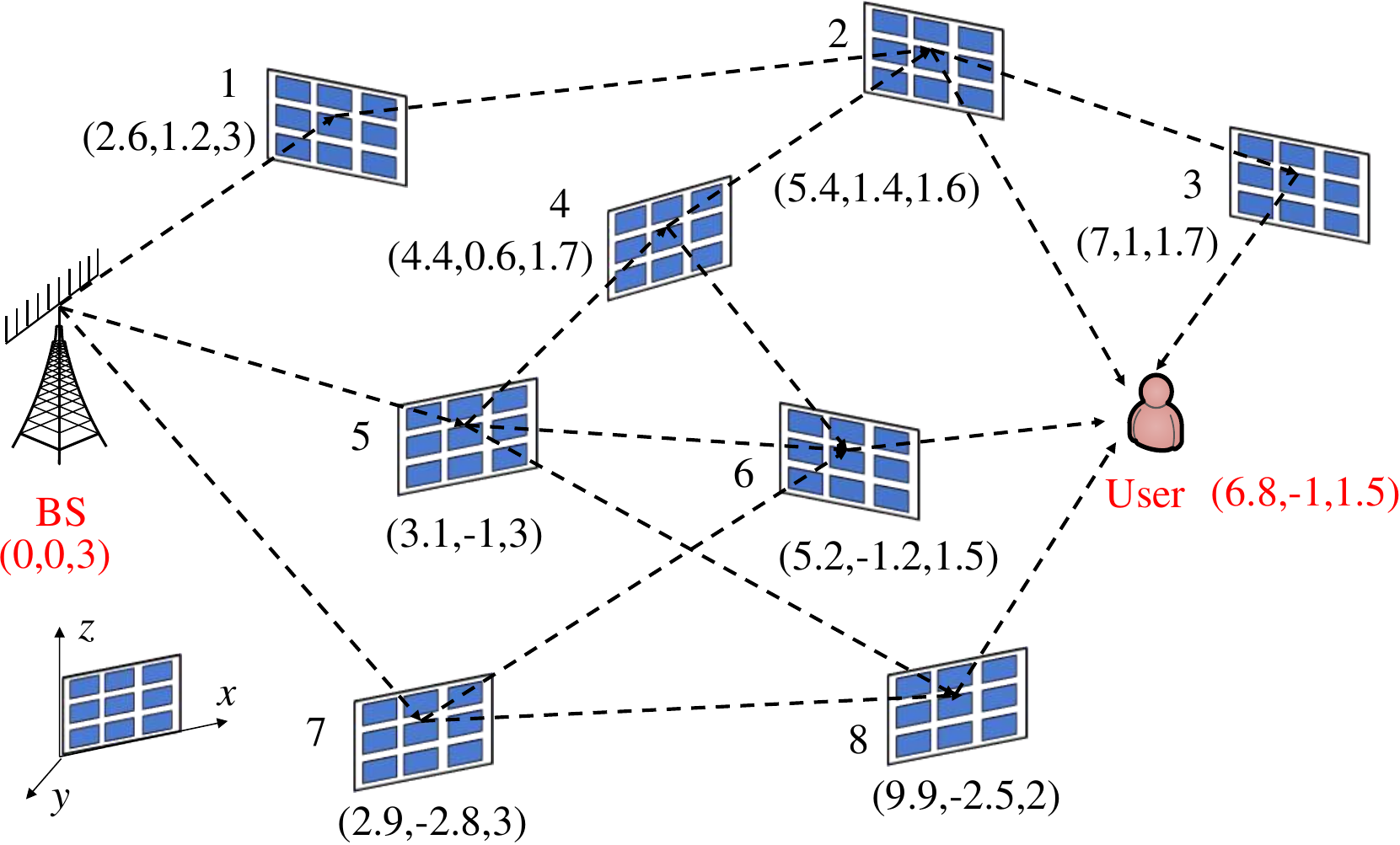}\label{single3D}}
	\subfigure[LoS graph.]{\includegraphics[width=0.35\textwidth]{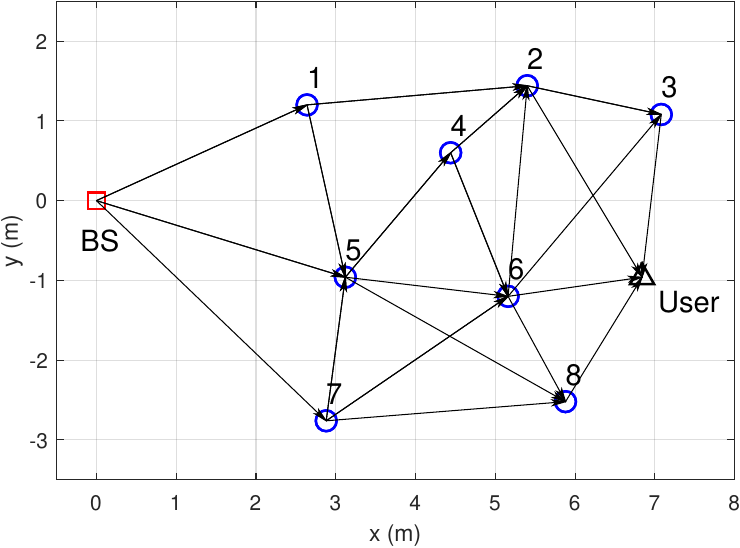}\label{LoSgraph}}
	\caption{Simulation setup for the single-user case.}\vspace{-12pt}
	\label{singlesetup}
\end{figure}

\begin{figure*}[!t]
	\centering
	\subfigure[Proposed scheme, $M_0=14$.]{\includegraphics[width=0.29\textwidth]{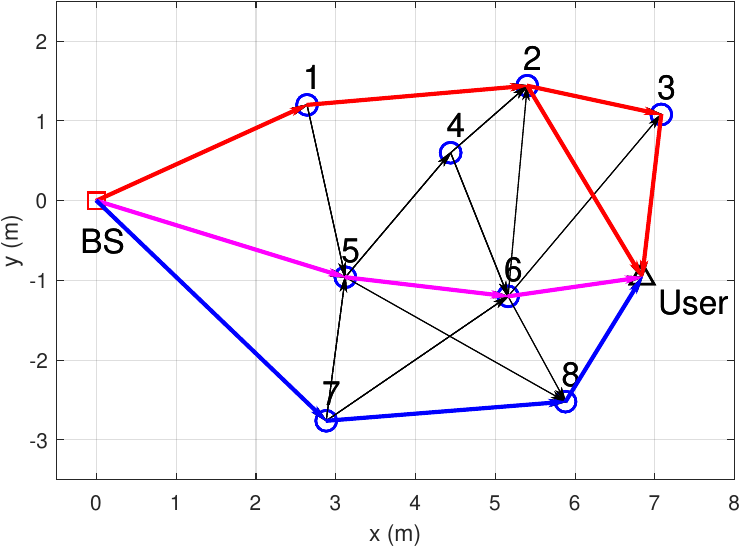}\label{Ses14}}
	\subfigure[Benchmark 1, $M_0=14$.]{\includegraphics[width=0.29\textwidth]{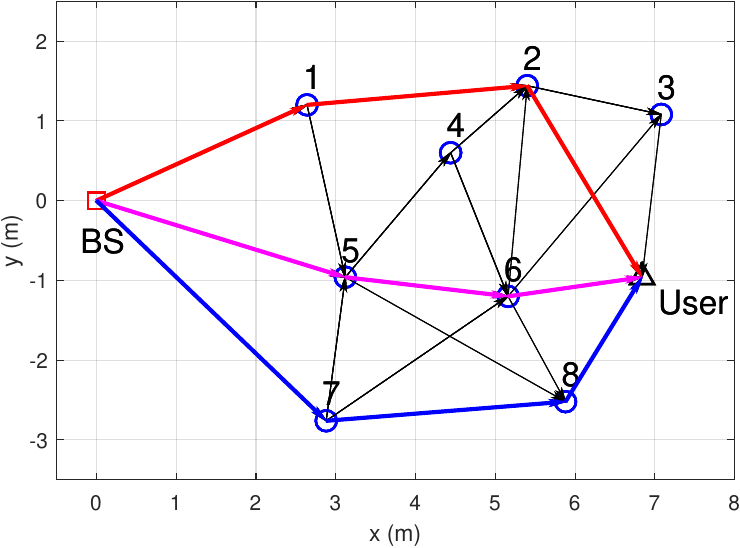}\label{Sms14}}
	\subfigure[Benchmark 2, $M_0=14$.]{\includegraphics[width=0.29\textwidth]{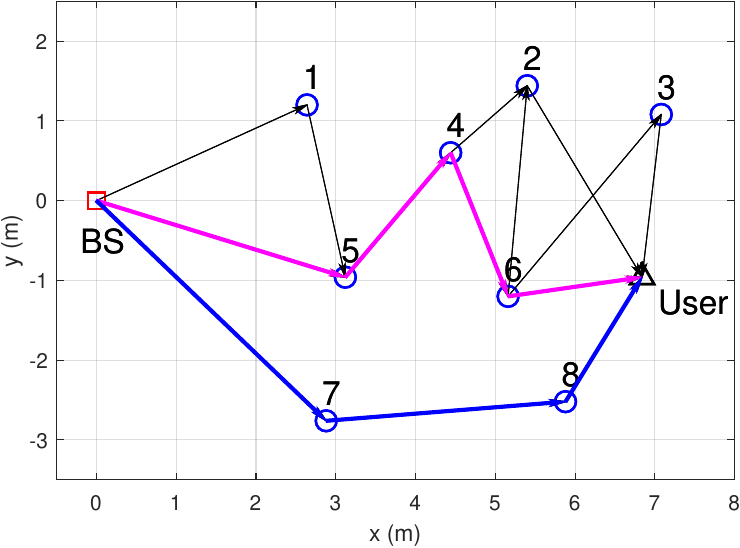}\label{Sreflectiononlya}}
	\subfigure[Proposed scheme, $M_0=16$.]{\includegraphics[width=0.29\textwidth]{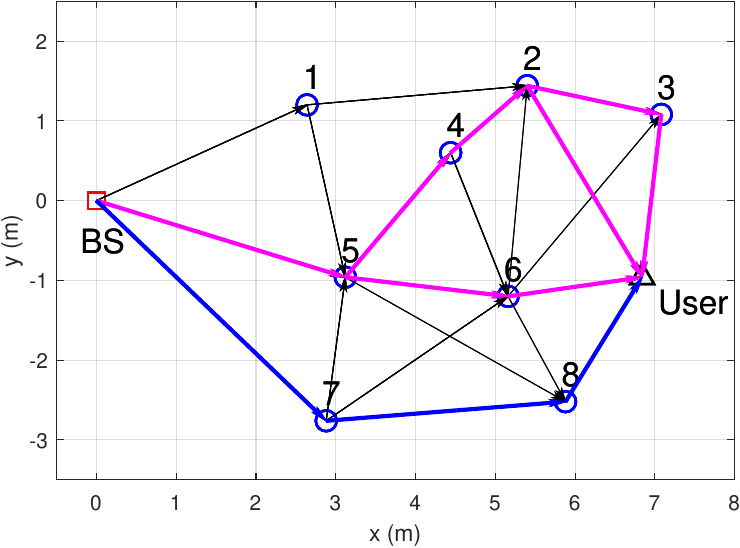}\label{Ses16}}
	\subfigure[Benchmark 1, $M_0=16$.]{\includegraphics[width=0.29\textwidth]{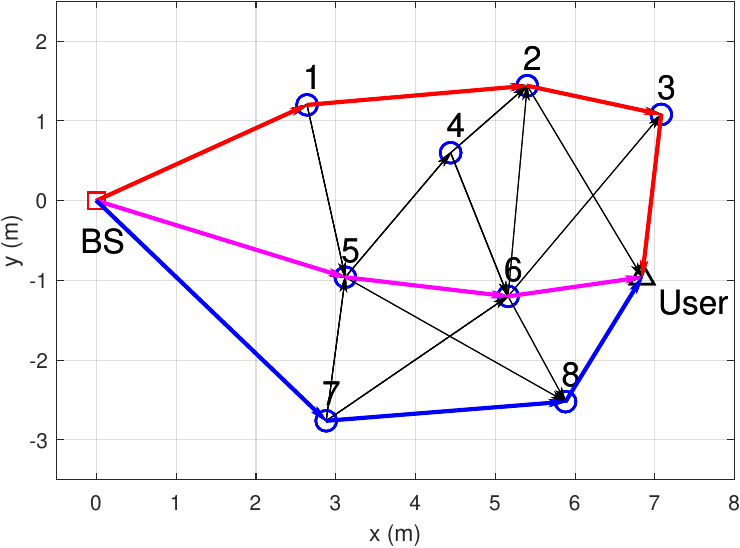}\label{Sms16}}
	\subfigure[Benchmark 2, $M_0=16$.]{\includegraphics[width=0.29\textwidth]{Sreflectiononly}\label{Sreflectiononlyb}}
	\caption{Selected paths by different schemes and $M_0$.}\label{Spaths}
\end{figure*}

Among all paths in $\cal S$, we aim to find the best set of paths that is feasible to (P3) and achieves the minimum objective value. To this end, we construct a path graph $G_p=(V_p,E_p)$ including all candidate paths in $\cal S$ (instead of $\mathcal{S}_k$), i.e., $V_p=\{v(P_k^{(s)})\vert k\in\mathcal{K}, s\in\{1,2,\cdots,S\}\}$, with $v(P_k^{(s)})$ denoting the vertex  corresponding to path $P_k^{(s)}$. Next, we add an edge between any two vertices in $V_p$, e.g., $v(P_k^{(s)})$ and $v(P_k^{(s')})$, if and only if the corresponding paths in $G_p$ satisfy the constraints in (P3). Instead of constructing $K$ graphs for the $K$ users respectively, we construct a single path graph for all $KS$ candidate paths to facilitate capturing the path selection constraints for each user and for different users at the same time. As a result, each clique in $G_p$ corresponds to a feasible solution to (P3) and it suffices to
find the maximal cliques in $G_p$. We apply the Bron-Kerbosch algorithm\cite{west2001introduction} to enumerate the maximal cliques in $G_p$, with the worst-case computational complexity in the order of $\mathcal O(3^{KS/3})$. Finally, we compare the objective value of (P3) achieved by each maximal clique in $G_p$ and select the best one as the output. Note that if the number of candidate paths $S$ is sufficiently large, the optimal solution should be included at th cost of higher complexity.

\begin{remark}
	Note that in addition to the proposed Yen's algorithm, some other heuristic algorithms, such as the Eppstein’s algorithm and K$^*$ search can also be used for path selection. However, the proposed algorithm enables a more flexible performance-cost trade-off by tuning the value of $S$, i.e., the number of candidate paths per user. As $S$ is sufficiently large, the proposed algorithm is ensured to achieve global optimality. As will be shown in Section \ref{numr} via simulation, a moderate number of $S$ suffices for our proposed algorithm to achieve a near-optimal performance.
\end{remark}
\begin{remark}
	An alternative approach for path selection is based on a mixed-integer programming formulation. However, this approach faces difficulties in our considered system. The reason is that a massive number of integer variables are required to express all the path selection constraints, which leads to high modeling and computational complexity.
\end{remark}

\begin{figure}[!t]
	\centering
	\includegraphics[width=0.35\textwidth]{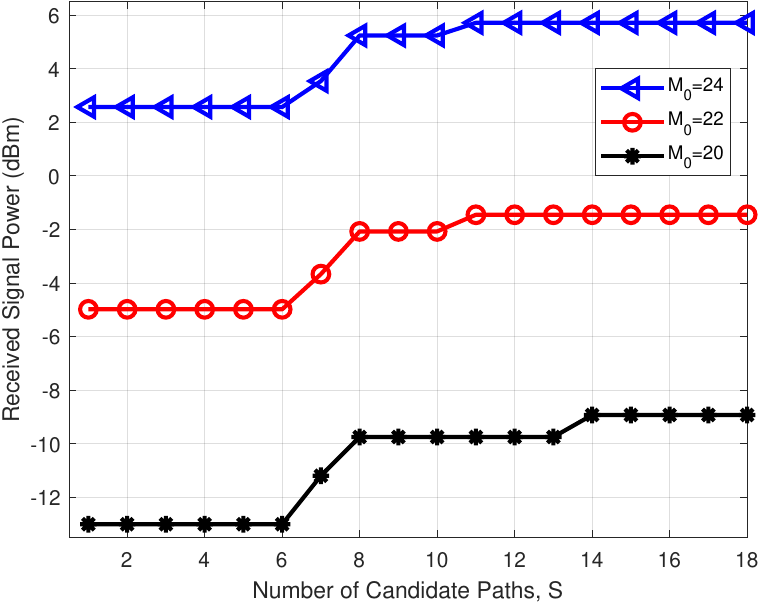}
	\caption{Received signal power versus the number of candidate paths, $S$.}\vspace{-6pt}
	\label{Senum}
\end{figure}
\section{Numerical Results}
\label{numr}
In this section, numerical results are provided to demonstrate the efficacy of our proposed STAR-RIS-enabled multi-beam routing. We set the number of the BS's antennas and the operating carrier frequency as $N_B=16$ and $f_c=5$ GHz, respectively, leading to $\gamma=(\frac{\lambda_c}{4\pi})^2=-46$ dB, where $\lambda_c=\frac{c}{f_c}$ denotes the wavelength with $c$ denoting the velocity of light. The number of elements in each dimension of the STAR-RISs is assumed to be identical to $M_{0}\triangleq\sqrt{M}$. For performance comparison, we consider two benchmark schemes with STAR-RISs in the MS mode (labeled as Benchmark 1) and conventional reflection-only RISs (labeled as Benchmark 2), respectively.

\vspace{-12pt}
\subsection{Single-User Case}
\label{Simulations}
As shown in Fig. \ref{singlesetup}, we consider $J=8$ STAR-RISs between the BS and user, and their 3D coordinates and formed LoS graph $G_L$ are shown in Fig. \ref{single3D} and Fig. \ref{LoSgraph}, respectively. For Benchmark 2, its corresponding LoS graph can be obtained by removing the edges $e_{1,2}$, $e_{2,3}$, $e_{5,6}$, $e_{5,8}$, $e_{6,8}$, $e_{7,5}$ and $e_{7,6}$ in Fig. \ref{LoSgraph}.

First, we study the optimized reflection paths by the proposed scheme and the two benchmark schemes for $M_0=$14 and 16 in Fig \ref{Spaths}. It is observed from Figs. \ref{Ses14}-\ref{Sreflectiononlya} that under $M_0=14$, the proposed scheme yields the maximum number (i.e., 4) of reflected paths from the BS to the user, while Benchmark 2 yields the minimum number (i.e., 2). This is expected as the proposed scheme can yield the highest LoS path diversity gain among the three schemes by leveraging the simultaneous signal reflection and transmission in the 360$^{\circ}$ full-space. The conventional reflection-only RISs can only achieve signal reflection in the 180$^{\circ}$ half-space, thus resulting the lowest LoS path diversity. It is also observed that the reflected paths in the proposed scheme include those in Benchmark 1 by further utilizing the transmission function of STAR-RIS 4. However, this does not hold for $M_0=16$, as observed from Figs. \ref{Ses16}-\ref{Sreflectiononlyb}. Moreover, it is also observed that the selected reflected paths by all considered schemes under $M_0=16$ generally go through more hops of STAR-RISs compared to $M_0=14$, so as to better exploit the more significant CPB gain with increasing $M_0$.
\begin{figure}[!t]
	\centering
	\includegraphics[width=0.35\textwidth]{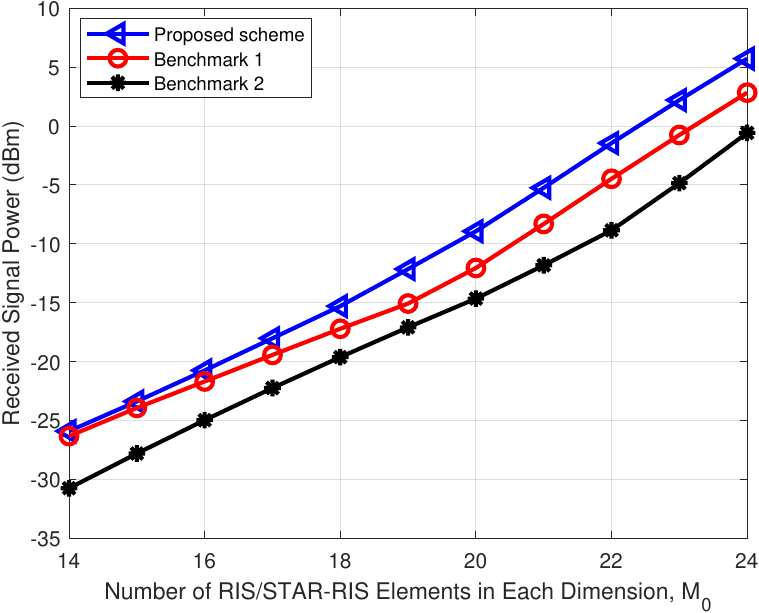}
	\caption{Received signal power versus the number of RIS/STAR-RIS elements in each dimension, $M_0$.}\vspace{-6pt}
	\label{SPvsM}
\end{figure}

Next, in Fig. \ref{Senum}, we study the received signal power at the user versus the number of candidate paths $S$ under different values of $M_0$. It is observed that the received signal power at the user is non-increasing with $S$ thanks to the enlarged solution set for path selection. However, the received signal power cannot be further increased by increasing $S$ when $S\ge10$, 8 and 9 under $M_0=$ 20, 22, and 24, respectively, which implies that the optimal path selection is likely to be achieved by the proposed algorithm under a relatively small $S$.

In Fig. \ref{SPvsM}, we study the received signal power at the user versus $M_0$ by different schemes. It is observed that our proposed scheme outperforms the two benchmarks over the whole range of $M_0$ considered. Particularly, the performance gain over Benchmark 1 is observed to become more significant for a larger $M_0$ and can approach 5 dB when $M_0=24$. The possible reason is that each reflected path may go through more STAR-RISs with increasing $M_0$, thus imposing more stringent constraints on the two benchmarks for selecting node-disjoint paths. In contrast, the proposed scheme allows for the selection of coupled paths and thus is less affected by increasing $M_0$.

\begin{figure}[!t]
	\centering
	\subfigure[3D plot.]{\includegraphics[width=0.35\textwidth]{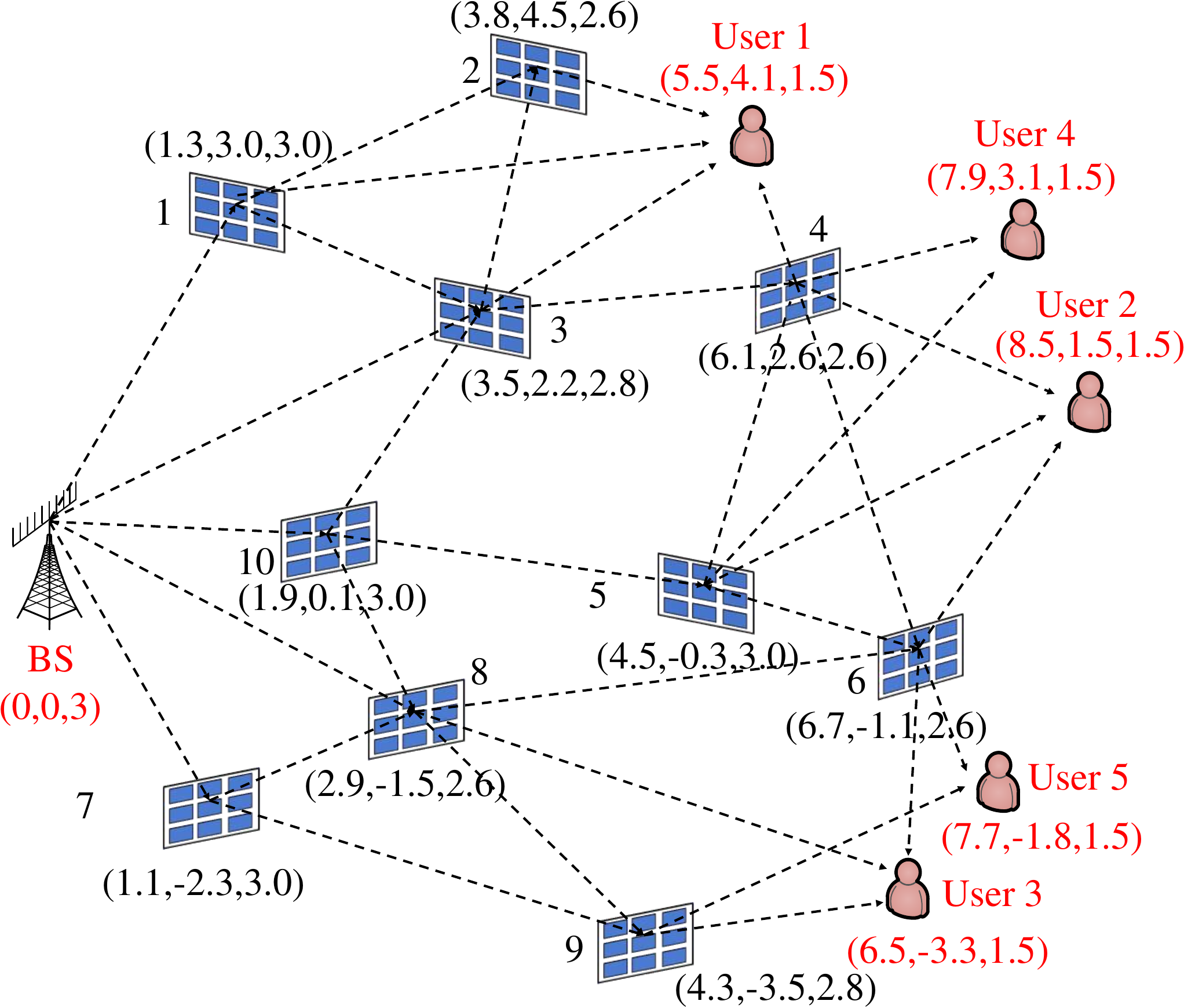}\label{M3D}}
	\subfigure[LoS graph.]{\includegraphics[width=0.35\textwidth]{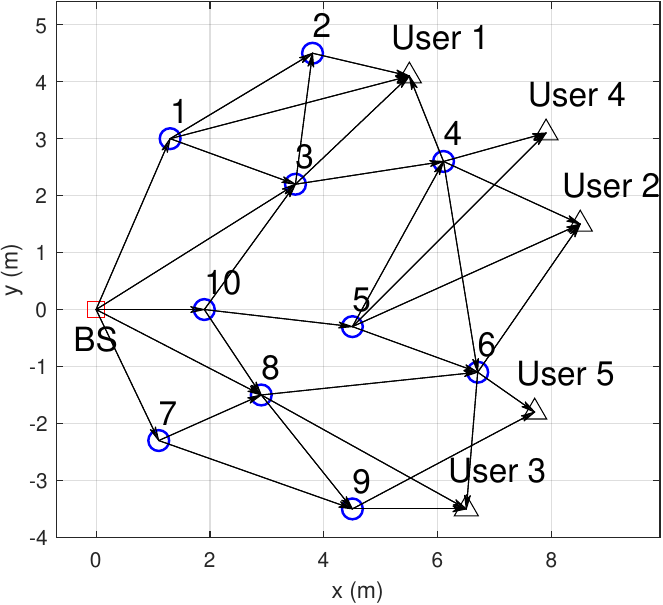}\label{MLoSgraph}}
	\caption{Simulation setup for the multi-user case.}\vspace{-12pt}
	\label{Msetup}
\end{figure}   

\subsection{Multi-User Case}
\label{MSimulations}
In this subsection, we consider maximally five users with the aid of $J=10$ STAR-RISs, and their 3D coordinates are shown in Fig. \ref{M3D}. The associated LoS graph is shown in Fig. \ref{MLoSgraph}. Note that more users can be served by the BS via multi-group multicasting or proper user scheduling over time with dynamic beam routing [18]. In the sequel of this subsection, we assume that users 1 to $i$  are assigned for communications when $K=i$.

\begin{figure*}[!t]
	\centering
	\subfigure[Proposed scheme, $M_0=14$.]{\includegraphics[width=0.29\textwidth]{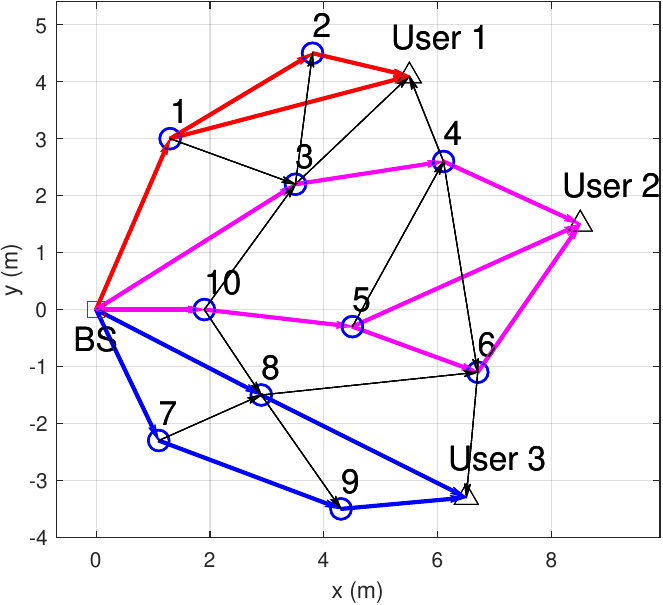}\label{Mes14}}
	\subfigure[Benchmark 1, $M_0=14$.]{\includegraphics[width=0.29\textwidth]{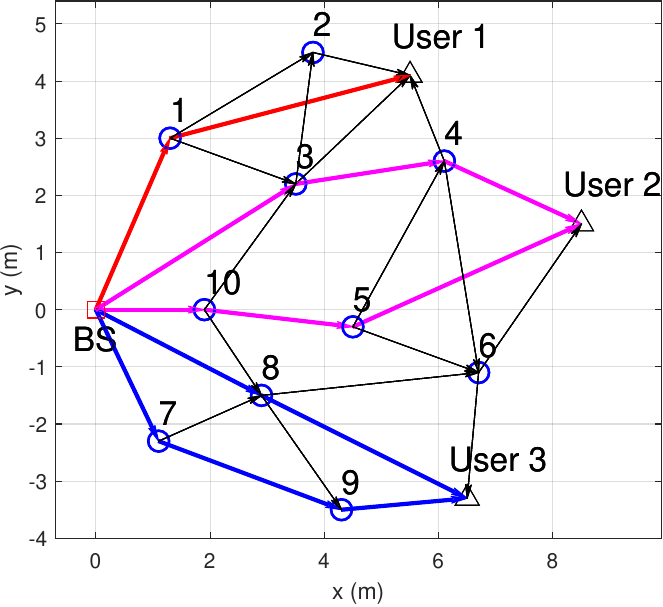}\label{Mms14}}
	\subfigure[Benchmark 2, $M_0=14$.]{\includegraphics[width=0.29\textwidth]{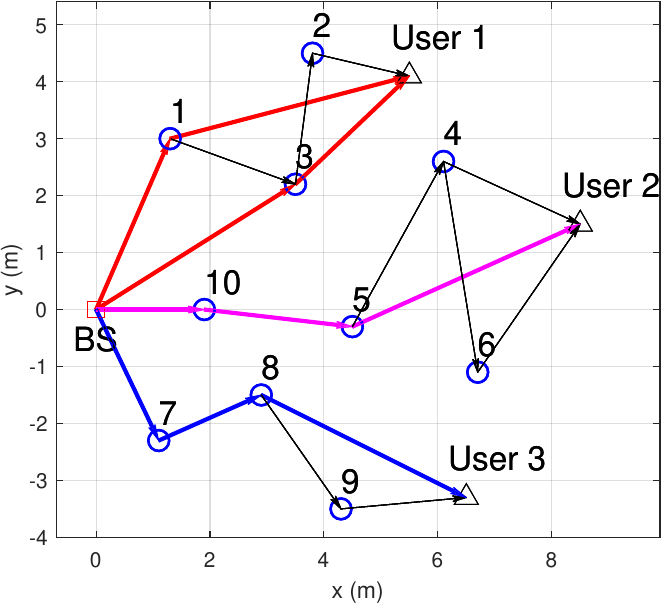}\label{Mr14}}
	\subfigure[Proposed scheme, $M_0=20$.]{\includegraphics[width=0.29\textwidth]{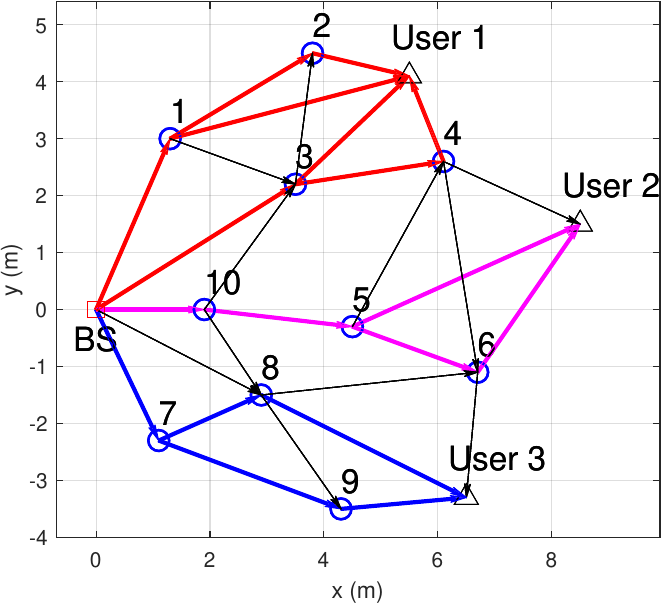}\label{Mes20}}
	\subfigure[Benchmark 1, $M_0=20$.]{\includegraphics[width=0.29\textwidth]{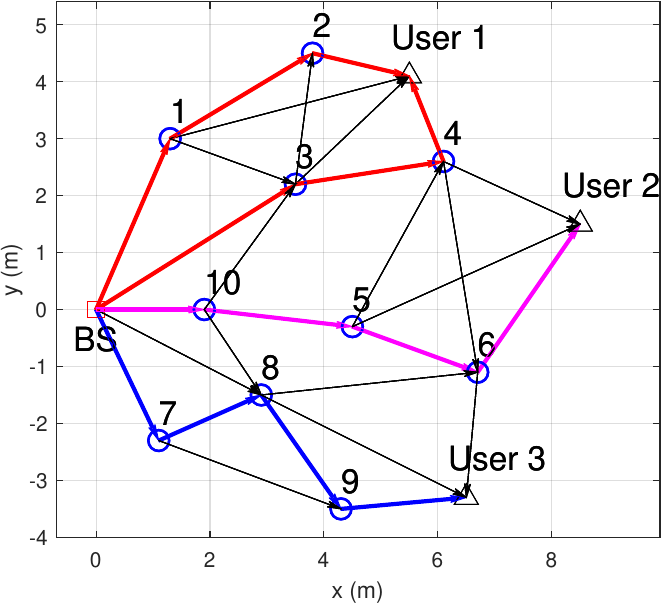}\label{Mms20}}
	\subfigure[Benchmark 2, $M_0=20$.]{\includegraphics[width=0.29\textwidth]{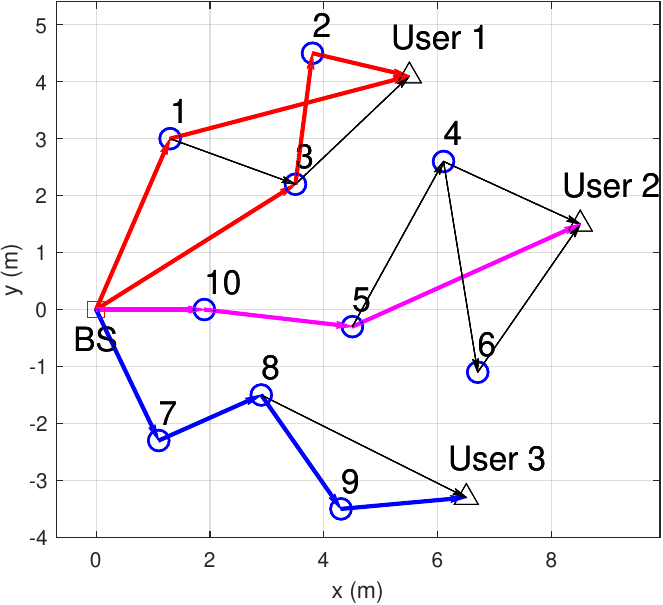}\label{Mr20}}
	\caption{Selected paths for multiple users by different schemes and $M_0$.}\label{Mpaths}\vspace{-12pt}
\end{figure*}

In Fig. \ref{Mpaths}, we show the selected paths by the proposed scheme and the two benchmark schemes for $M_0=14$ and 20 and $K=3$. It is observed that similar to the single-user case, the proposed scheme yields the largest number of paths for each user among the considered schemes thanks to its highest LoS path diversity. In contrast, Benchmark 2 yields the minimum number of paths due to its lowest LoS path diversity with half-space reflection constraint. In particular, in the MS mode, it is observed from Figs. \ref{Mms14} and \ref{Mms20} that the number of selected paths for users 1 and 2 may affect each other. As a result, one of users 1 and 2 is served by two paths, while the other is only served by a single path. Unlike the MS mode, the proposed algorithm can assign at least 2 paths for each user, thereby greatly improving their communication performance. It is also observed that when $M_0$ increases, the selected paths go through more hops of STAR-RISs to enjoy the higher CPB gain, which is consistent with the observation made in Fig. \ref{MPvsM}. Furthermore, it is observed that the selected reflected paths by the proposed scheme ensure separation among all users, which manifests the effectiveness of our proposed clique-based algorithm.

Fig. \ref{MPvsS} shows the max-min received signal power versus the number of candidate paths selected for each user, $S$, with $K=3$. It is observed that the max-min received signal power cannot be further improved by increasing the number of candidate paths when $S\geq5$, 9 and 6 under $M_0=14$, 18 and 22, respectively, implying that the optimal solution may have been achieved by the proposed algorithm with a relatively small $S$ even in the multi-user case. This thus ensures the scalability of our proposed beam-routing scheme with increasing the number of users.
\begin{figure}[!t]
	\centering
	\includegraphics[width=0.35\textwidth]{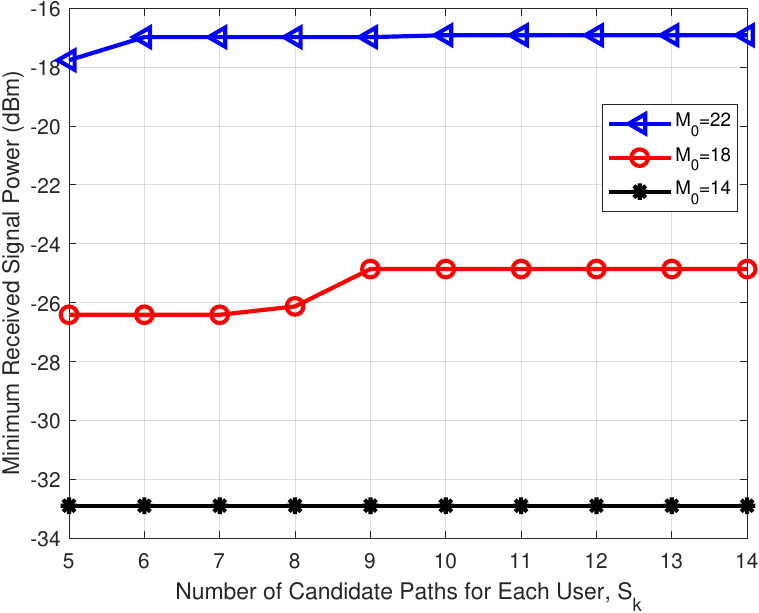}
	\caption{Max-min received signal power versus the number of candidate paths.}\vspace{-6pt}
	\label{MPvsS}
\end{figure}

In Fig. \ref{MPvsM}, we show the max-min received signal power among $K=3$ users versus the number of RIS/STAR-RIS elements in each dimension $M_0$. It is observed that the performance by all schemes increases with $M_0$ thanks to the increased CPB gain. Moreover, our proposed scheme outperforms two benchmarks over the whole range of $M_0$ considered. In particular, Benchmark 1 with the MS mode is observed to even achieve the same performance as Benchmark 2 with reflection-only RISs as $M_0$ is small. In contrast, the proposed scheme can achieve 2-3 dB higher max-min received signal power compared to the two benchmark schemes for all $M_0$.
\begin{figure}[!t]
	\centering
	\includegraphics[width=0.35\textwidth]{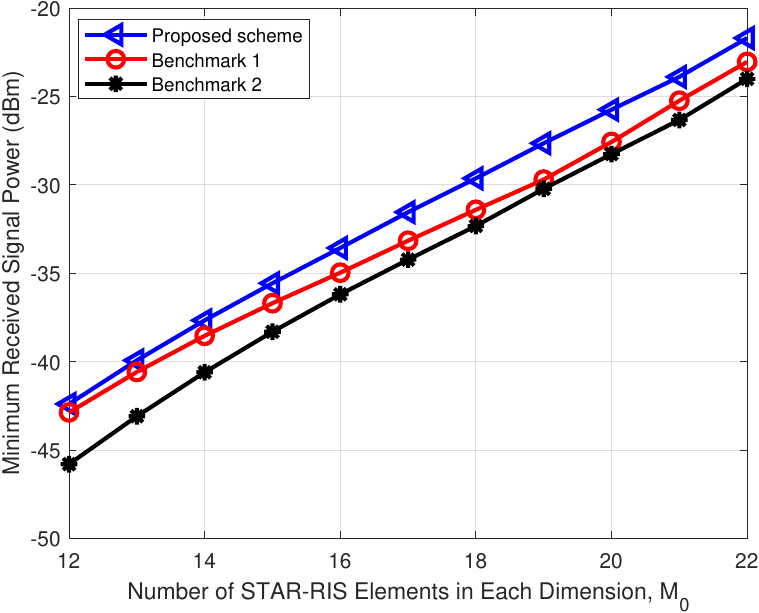}
	\caption{Max-min received signal power versus the number of RIS/STAR-RIS elements in each dimension, $M_0$.}\vspace{-6pt}
	\label{MPvsM}
\end{figure}

Finally, in Fig. \ref{MPvsK}, we show the max-min received signal power versus the number of users, $K$. It is first observed that increasing the number of users decreases the max-min received signal power by all schemes, as expected. It is also observed that the proposed scheme achieves better performance than the two benchmarks for all considered user numbers, as more LoS signal paths can be created by the proposed scheme to serve more users at the same time. In particular, Benchmark 2 becomes infeasible to serve more than three users simultaneously due to the half-space reflection of conventional RISs, which dramatically limits the number of LoS signal paths in the system. Moreover, the performance gap between the proposed method and Benchmark 1 gradually decreases as the number of users increases. This is because, with more users, a greater number of STAR-RISs are constrained to operate in either the reflective or transmissive mode (rather than both) to satisfy the path disjointness constraints.

\section{Conclusion}\label{conclusion}
In this paper, we proposed a multi-STAR-RIS enabled multi-path beam routing scheme for a multi-user communication system. First, we focused on a simplified single-user system setup and derived the optimal active beamforming and power allocation at the BS, as well as the optimal amplitude and phase-shifts for both transmission and reflection in closed form for a given path selection. The results unveil that the maximum received signal power at the user can be expressed as the sum of the maximum channel power gains of the selected paths as if they were disjoint. A clique-based algorithm was then proposed to obtain a near-optimal solution for the path selection problem. Next, we extended our proposed methods to the multicast scenario subject to additional node-disjoint constraints on the selected paths for different users. Numerical results demonstrate that the proposed algorithm can achieve near-optimal performance and exploiting STAR-RISs for beam routing can achieve a considerably better performance compared to conventional reflection-only RISs by reaping more pronounced LoS path diversity gain. This paper can be extended in several directions for future work. For example, it is worthy of studying a hybrid active and passive STAR-RIS system to further compensate for the cascaded path losses due to multi-hop signal transmission/reflection. Moreover, the hybrid deployment strategy for multiple RISs and STAR-RISs is also an interesting and practically relevant problem that deserves in-depth investigation. Last but not least, it is worthy of studying practical routing designs accounting for near-field channel effects especially when the sizes of STAR-RISs are large, as well as more general node-joint multi-path beam routing in the multi-user case.
\vspace{-12pt}
\begin{figure}[!t]
	\centering
	\includegraphics[width=0.35\textwidth]{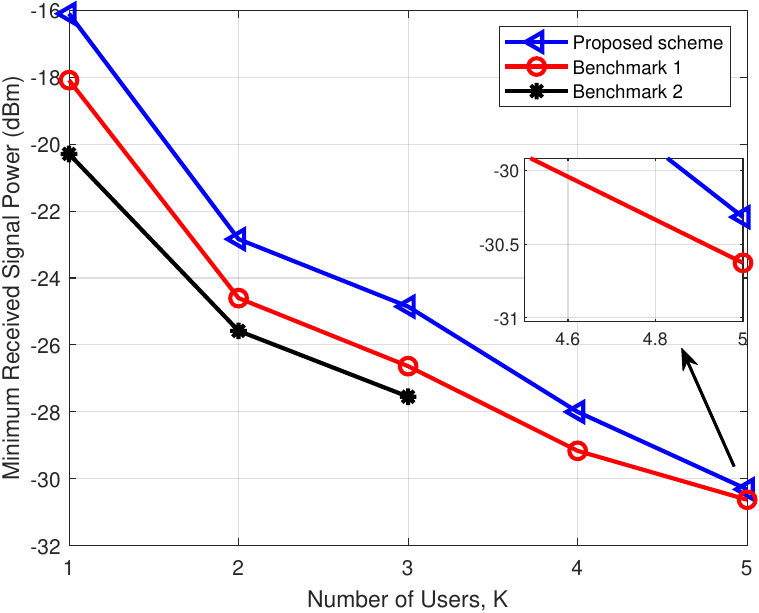}
	\caption{{Max-min received signal power versus the number of users, $K$.}}\vspace{-6pt}
	\label{MPvsK}
\end{figure}

{\appendix[Proof of Proposition \ref{Prop1}]\label{Proof1}
To derive the optimal reflected and transmitted amplitude of STAR-RIS $j$, we first denote the product of the amplitude of STAR-RIS $j$'s subsequent nodes over each reflected/transmitted path $c_l^{(j,\mu)}$ as $B_{l,\text{S}}^{(j,\mu)}$, $\mu\in\{R,T\}$. Note that only when the path going through STAR-RIS $j$ is further split into more paths by its subsequent nodes, we have $L_{j,\mu}\geq2$. Evidently, for STAR-RIS $j$ with $L_{j,\mu}=1$, we have $B_{1,\text{S}}^{(j,\mu)}=1$ because only one reflected/transmitted path goes through it without passive beam splitting. Note that if there is no subsequent node after STAR-RIS $j$, $B_{l,\text{S}}^{(j,\mu)}$ can be set to unity. For example, in Fig. \ref{esms}, we have $B_{1,\text{S}}^{(1,T)}=\beta_{2}^{(T)}$, $B_{2,\text{S}}^{(1,T)}=\beta_{2}^{(R)}$, $B_{2,\text{S}}^{(1,R)}=1$ for STAR-RIS 1 and  $B_{1,\text{S}}^{(2,T)}=B_{1,\text{S}}^{(2,R)}=1$ for STAR-RIS 2.
	
As shown in Fig. \ref{proof1}, we assume that the $q$-th active beam is first split at STAR-RIS $i$. Since the previous nodes only reflect/transmit the signal, the product of their amplitudes should be unity, and the sum of channel power gains over all paths associated with the $q$-th active beam is expressed as
\begin{align}
			P_{J+1}^{(q)}&=\alpha_q\left( \sqrt{\beta_i^{(R)}}\sum_{l=1}^{L_{i,R}}\sqrt{B_{l,\text{S}}^{(i,R)}\hat{F}_{0,J+1}(c_l^{(i,R)})}\right.\nonumber\\
			&\qquad\qquad\left.+\sqrt{\beta_i^{(T)}}\sum_{l=1}^{L_{i,T}}\sqrt{B_{l,\text{S}}^{(i,T)}\hat{F}_{0,J+1}(c_l^{(i,T)})}\right)^2\nonumber\\
			&\leq\alpha_q\left( \beta_i^{(R)}+\beta_i^{(T)} \right) \left( \left(\sum_{l=1}^{L_{i,R}}\sqrt{B_{l,\text{S}}^{(i,R)}\hat{F}_{0,J+1}(c_l^{(i,R)})}\right)^2\right.\nonumber\\
			&\qquad\qquad\left.+\left(\sum_{l=1}^{L_{i,T}}\sqrt{B_{l,\text{S}}^{(i,T)}\hat{F}_{0,J+1}(c_l^{(i,T)})}\right)^2\right)\nonumber\\
			&=\alpha_{q}\left(P_{J+1}^{(i,R)}+P_{J+1}^{(i,T)}\right)\triangleq	\tilde{P}_{J+1}^{(q)},\label{activeq}
\end{align}
where $P_{J+1}^{(i,\mu)}=\left( \sum_{l=1}^{L_{i,\mu}}\sqrt{B_{l,\text{S}}^{(i,\mu)}\hat{F}_{0,J+1}(c_l^{(i,\mu)})}\right)^2$ and the inequality is due to the Cauchy-Schwarz inequality. Note that each path associated with the $q$-th active beam corresponds to one reflected/transmitted path of STAR-RIS $i$. As such, \eqref{proposition1pw} can be proved if we can prove
\begin{equation}
		\begin{aligned}
			P_{J+1}^{(j,\mu)}\leq\sum_{l=1}^{L_{j,\mu}}\hat{F}_{0,J+1}(c_l^{(j,\mu)})\triangleq\tilde{P}_{J+1}^{(j,\mu)},\: \forall j\in\tilde{\Lambda},\label{mathintro}
		\end{aligned}
\end{equation}
where $L_{j,\mu}\geq1$. Next, we prove \eqref{mathintro} for each selected STAR-RIS $j$ via induction. First, we validate \eqref{mathintro} in the case shown in \eqref{mathintro} with $L_{j,\mu}=1$ and 2. Then, we assume that \eqref{mathintro} always holds when $L_{j,\mu}\leq L_0$, with $L_0\geq 3$. Finally, we prove that \eqref{mathintro} still holds when $L_{j,\mu}=L_0+1$. The details are as follows.
	
First, it can be observed that the example shown in Fig. \ref{esms} can satisfy \eqref{mathintro}, i.e., $P_{J+1}^{(1,T)}=\hat{F}_{0,J+1}(\Omega_{1,1})+\hat{F}_{0,J+1}(\Omega_{1,2})$ with $L_{1,T}=2$, $P_{J+1}^{(1,R)}=P_{J+1}^{(4,R)}=\hat{F}_{0,J+1}(\Omega_{1,3})$ with $L_{1,R}=1$ and $L_{4,R}=1$, $P_{J+1}^{(2,T)}=P_{J+1}^{(3,R)}=\hat{F}_{0,J+1}(\Omega_{1,1})$ with $L_{2,T}=1$ and $L_{3,R}=1$, $P_{J+1}^{(2,R)}=\hat{F}_{0,J+1}(\Omega_{1,2})$ with $L_{2,R}=1$, as well as $P_{J+1}^{(5,T)}=P_{J+1}^{(6,R)}=\hat{F}_{0,J+1}(\Omega_{2,1})$ with $L_{5,T}=1$ and $L_{6,R}=1$. Next, we assume that for any STAR-RIS $j$ with $L_{j,\mu}\leq L_0$, \eqref{mathintro} always holds. Then, we consider STAR-RIS $j$ with $L_{j,\mu}= L_0+1$.
	
\begin{figure}[!t]
		\centering
		\includegraphics[width=3.5in]{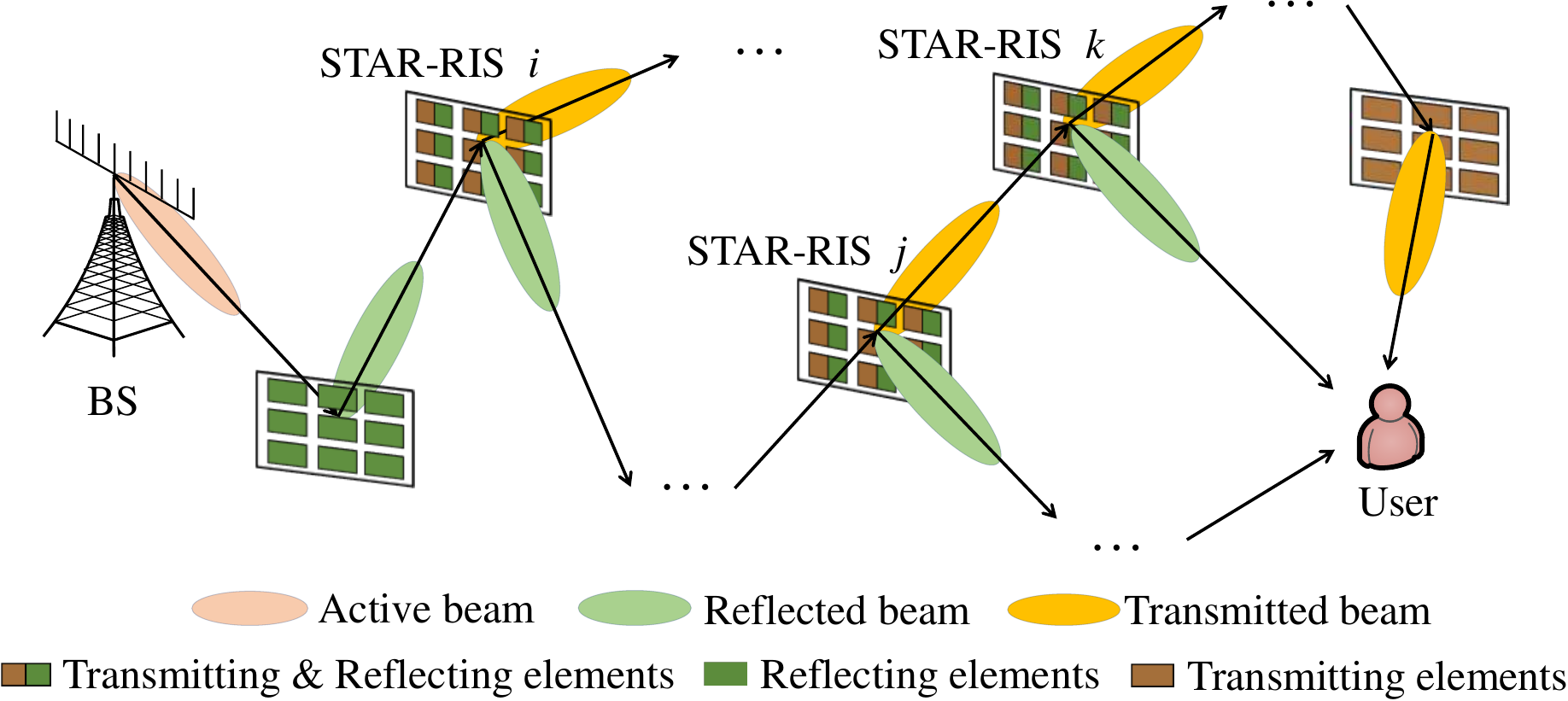}
		\caption{An illustrative example of Proposition 1.}
		\label{proof1}\vspace{-12pt}
\end{figure}
	
For STAR-RIS $j$ with $L_{j,\mu}= L_0+1$, assuming that node $k$ is its first subsequent node, $k\in\mathcal{D}$, we have
\begin{align}
		P_{J+1}^{(j,\mu)}=&\left( \sum_{l=1}^{L_{j,\mu}}\sqrt{B_{l,\text{S}}^{(j,\mu)}\hat{F}_{0,J+1}(c_l^{(j,\mu)})}\right)^2\nonumber\\
		=&\left( \sum_{l=1}^{L_{k,R}}\sqrt{\beta_k^{(R)}B_{l,\text{S}}^{(k,R)}\hat{F}_{0,J+1}(c_l^{(k,R)})}\right.\nonumber\\
		&\qquad\qquad\left.+\sum_{l=1}^{L_{k,T}}\sqrt{\beta_k^{(T)}B_{l,\text{S}}^{(k,T)}\hat{F}_{0,J+1}(c_l^{(k,T)})}\right)^2\nonumber\\
		\leq &\left( \beta_k^{(R)}+\beta_k^{(T)} \right) \left( \left(\sum_{l=1}^{L_{k,R}}\sqrt{B_{l,\text{S}}^{(k,R)}\hat{F}_{0,J+1}(c_l^{(k,R)})}\right)^2\right.\nonumber\\
		&\qquad\qquad\left.+\left( \sum_{l=1}^{L_{k,T}}\sqrt{B_{l,\text{S}}^{(k,T)}\hat{F}_{0,J+1}(c_l^{(k,T)})}\right)^2\right)\nonumber\\
		\leq&\tilde{P}_{J+1}^{(k,R)}+\tilde{P}_{J+1}^{(k,T)}=\tilde{P}_{J+1}^{(j,\mu)}.\label{Pj}
\end{align}
It is evident that we have $L_{j,\mu}=L_{k,R}+L_{k,T}$ with $L_{k,R}\leq L_0$ and $L_{k,T}\leq L_0$. Accordingly, \eqref{Pj} can be written as 
\begin{align}
			\tilde{P}_{J+1}^{(j,\mu)}=&\sum_{l=1}^{L_{j,R}}\hat{F}_{0,J+1}(c_l^{k,R})+\sum_{l=1}^{L_{k,T}}\hat{F}_{0,J+1}(c_l^{k,T})\nonumber\\
			=&\sum_{l=1}^{L_{j,\mu}}\hat{F}_{0,J+1}(c_l^{j,\mu}),
\end{align}
where the last equality is because each path going through the transmission/reflection surface of STAR-RIS $j$ (i.e. $c_l^{(j,\mu)}, l\in\{1,2,\cdots, L_{j,\mu}\}$) corresponds to a transmitted/reflected path of STAR-RIS $k$ (i.e. $c_l^{(k,\mu)}, l\in\{1,2,\cdots, L_{k,\mu}\}$). Hence, \eqref{mathintro} is proved. Moreover, it can be shown that by setting
\begin{align}
			\beta_{k}^{(\mu)}=&\frac{\tilde{P}_{J+1}^{(k,\mu)}}{\tilde{P}_{J+1}^{(k,R)}+\tilde{P}_{J+1}^{(k,T)}}\nonumber\\
			=&\frac{\sum_{l=1}^{L_{k,\mu}}\hat F(c^{(k,\mu)}_l)}{\sum_{l=1}^{L_{k,T}}\hat F(c^{(k,T)}_l)+\sum_{l=1}^{L_{k,R}}\hat F(c^{(k,R)}_l)},\quad \mu\in\{R,T\},\nonumber
\end{align}
the equality in \eqref{Pj} can always hold. Given \eqref{mathintro}, for the $q$-th active beam, \eqref{activeq} can be written as
\begin{equation}
		\begin{split}
			\tilde{P}_{J+1}^{(q)}=&\tilde{P}_{J+1}^{(i,R)}+\tilde{P}_{J+1}^{(i,T)}\\=&\sum_{l=1}^{L_{i,R}}\hat{F}_{0,J+1}(c_l^{(i,R)})+ \sum_{l=1}^{L_{i,T}}\hat{F}_{0,J+1}(c_l^{(i,T)})\\=&\sum_{p=1}^{P_q}\hat{F}_{0,J+1}(\Omega_{q,p}).
		\end{split}
\end{equation}
Then, considering all $Q$ active beams, we have $\tilde{P}_{J+1}=\sum_{q=1}^{Q}\tilde{P}_{J+1}^{(q)}=\sum\nolimits_{q=1}^Q\sum_{p=1}^{P_q}\hat{F}_{0,J+1}(\Omega_{q,p})$. This thus completes the proof.}

\bibliographystyle{IEEEtran}
\bibliography{IEEEabrv,reference}

\begin{thebibliography}{10}
\providecommand{\url}[1]{#1}
\csname url@samestyle\endcsname
\providecommand{\newblock}{\relax}
\providecommand{\bibinfo}[2]{#2}
\providecommand{\BIBentrySTDinterwordspacing}{\spaceskip=0pt\relax}
\providecommand{\BIBentryALTinterwordstretchfactor}{4}
\providecommand{\BIBentryALTinterwordspacing}{\spaceskip=\fontdimen2\font plus
\BIBentryALTinterwordstretchfactor\fontdimen3\font minus
  \fontdimen4\font\relax}
\providecommand{\BIBforeignlanguage}[2]{{%
\expandafter\ifx\csname l@#1\endcsname\relax
\typeout{** WARNING: IEEEtran.bst: No hyphenation pattern has been}%
\typeout{** loaded for the language `#1'. Using the pattern for}%
\typeout{** the default language instead.}%
\else
\language=\csname l@#1\endcsname
\fi
#2}}
\providecommand{\BIBdecl}{\relax}
\BIBdecl

\bibitem{anstarris}
B.~An and W.~Mei, ``{STAR-RIS}-enabled multi-path beam routing with passive
  beam splitting,'' in \emph{Proc. IEEE Glob. Commun. Conf.}, Cape Town, South
  Africa, Dec. 2024, pp. 3553--3558.

\bibitem{9326394}
Q.~Wu, S.~Zhang, B.~Zheng, C.~You, and R.~Zhang, ``Intelligent reflecting
  surface-aided wireless communications: A tutorial,'' \emph{{IEEE} Trans.
  Commun.}, vol.~69, no.~5, pp. 3313--3351, May 2021.

\bibitem{9424177}
Y.~Liu, X.~Liu, X.~Mu, T.~Hou, J.~Xu, M.~Di~Renzo, and N.~Al-Dhahir,
  ``Reconfigurable intelligent surfaces: Principles and opportunities,''
  \emph{{IEEE} Commun. Surveys Tuts.}, vol.~23, no.~3, pp. 1546--1577, 3rd
  Quart. 2021.

\bibitem{renzo2019smart}
M.~D. Renzo \emph{et~al.}, ``Smart radio environments empowered by
  reconfigurable {AI} meta-surfaces: An idea whose time has come,''
  \emph{EURASIP J. Wireless Commun. Netw.}, vol. 2019, no.~1, pp. 1--20, May
  2019.

\bibitem{8910627}
Q.~Wu and R.~Zhang, ``Towards smart and reconfigurable environment: Intelligent
  reflecting surface aided wireless network,'' \emph{{IEEE} Commun. Mag.},
  vol.~58, no.~1, pp. 106--112, 2020.

\bibitem{basar2019wireless}
E.~Basar, M.~Di~Renzo, J.~De~Rosny, M.~Debbah, M.-S. Alouini, and R.~Zhang,
  ``Wireless communications through reconfigurable intelligent surfaces,''
  \emph{{IEEE} Access}, vol.~7, pp. 116\,753--116\,773, 2019.

\bibitem{di2020smart}
M.~Di~Renzo, A.~Zappone, M.~Debbah, M.-S. Alouini, C.~Yuen, J.~De~Rosny, and
  S.~Tretyakov, ``Smart radio environments empowered by reconfigurable
  intelligent surfaces: How it works, state of research, and the road ahead,''
  \emph{{IEEE} J. Sel. Areas Commun.}, vol.~38, no.~11, pp. 2450--2525, 2020.

\bibitem{yuan2021reconfigurable}
X.~Yuan, Y.-J.~A. Zhang, Y.~Shi, W.~Yan, and H.~Liu,
  ``Reconfigurable-intelligent-surface empowered wireless communications:
  Challenges and opportunities,'' \emph{{IEEE} Wireless Commun.}, vol.~28,
  no.~2, pp. 136--143, 2021.

\bibitem{pan2022overview}
C.~Pan, G.~Zhou, K.~Zhi, S.~Hong, T.~Wu, Y.~Pan, H.~Ren, M.~Di~Renzo, A.~L.
  Swindlehurst, R.~Zhang \emph{et~al.}, ``An overview of signal processing
  techniques for {RIS/IRS}-aided wireless systems,'' \emph{{IEEE} J. Sel.
  Topics Signal Process.}, vol.~16, no.~5, pp. 883--917, 2022.

\bibitem{zheng2022survey}
B.~Zheng, C.~You, W.~Mei, and R.~Zhang, ``A survey on channel estimation and
  practical passive beamforming design for intelligent reflecting surface aided
  wireless communications,'' \emph{{IEEE} Commun. Surveys Tuts.}, vol.~24,
  no.~2, pp. 1035--1071, 2022.

\bibitem{pathloss}
W.~Mei, B.~Zheng, C.~You, and R.~Zhang, ``Intelligent reflecting surface-aided
  wireless networks: From single-reflection to multireflection design and
  optimization,'' \emph{Proc. {IEEE}}, vol. 110, no.~9, pp. 1380--1400, 2022.

\bibitem{9410457}
C.~Huang, Z.~Yang, G.~C. Alexandropoulos, K.~Xiong, L.~Wei, C.~Yuen, Z.~Zhang,
  and M.~Debbah, ``Multi-hop {RIS}-empowered terahertz communications: A
  {DRL}-based hybrid beamforming design,'' \emph{{IEEE} J. Sel. Areas Commun.},
  vol.~39, no.~6, pp. 1663--1677, 2021.

\bibitem{mei2020cooperative}
W.~Mei and R.~Zhang, ``Cooperative beam routing for multi-{IRS} aided
  communication,'' \emph{{IEEE} Commun. Lett.}, vol.~10, no.~2, pp. 426--430,
  2020.

\bibitem{mei2021multi}
W.~Mei and R.~Zhang, ``Multi-beam multi-hop routing for intelligent reflecting
  surfaces aided massive {MIMO},'' \emph{{IEEE} Trans. Wireless Commun.},
  vol.~21, no.~3, pp. 1897--1912, 2021.

\bibitem{9829192}
X.~Ma, Y.~Fang, H.~Zhang, S.~Guo, and D.~Yuan, ``Cooperative beamforming design
  for multiple {RIS}-assisted communication systems,'' \emph{{IEEE} Trans.
  Wireless Commun.}, vol.~21, no.~12, pp. 10\,949--10\,963, 2022.

\bibitem{10589431}
Y.~Wu, T.~Zhao, H.~Hai, and E.~Bai, ``An index modulation system for
  multi-{IRS} aided communication,'' \emph{{IEEE} Commun. Lett.}, vol.~29,
  no.~1, pp. 6--10, 2025.

\bibitem{mei2022intelligent}
W.~Mei and R.~Zhang, ``Intelligent reflecting surface for multi-path beam
  routing with active/passive beam splitting and combining,'' \emph{{IEEE}
  Commun. Lett.}, vol.~26, no.~5, pp. 1165--1169, May 2022.

\bibitem{10643789}
W.~Mei, D.~Wang, Z.~Chen, and R.~Zhang, ``Joint beam routing and resource
  allocation optimization for multi-{IRS}-reflection wireless power transfer,''
  \emph{{IEEE} Trans. Wireless Commun.}, vol.~23, no.~11, pp. 16\,606--16\,620,
  2024.

\bibitem{zhang2022multi}
Y.~Zhang and C.~You, ``Multi-hop beam routing for hybrid active/passive {IRS}
  aided wireless communications,'' in \emph{Proc. IEEE Glob. Commun. Conf.},
  Rio de Janeiro, Brazi, Dec. 2022, pp. 3138--3143.

\bibitem{9900387}
A.~Bhowal and S.~Aïssa, ``{MIMO} device-to-device communications via
  cooperative dual-polarized intelligent surfaces,'' \emph{{IEEE} Wireless
  Commun. Lett.}, vol.~12, no.~2, pp. 202--206, 2023.

\bibitem{9800900}
D.~Tyrovolas, S.~A. Tegos, E.~C. Dimitriadou-Panidou, P.~D. Diamantoulakis,
  C.~K. Liaskos, and G.~K. Karagiannidis, ``Performance analysis of cascaded
  reconfigurable intelligent surface networks,'' \emph{{IEEE} Wireless Commun.
  Lett.}, vol.~11, no.~9, pp. 1855--1859, 2022.

\bibitem{10057425}
Z.~Zhakipov, K.~M. Rabie, X.~Li, and G.~Nauryzbayev, ``Accurate approximation
  to channel distributions of cascaded {RIS}-aided systems with phase errors
  over nakagami-m channels,'' \emph{{IEEE} Wireless Commun. Lett.}, vol.~12,
  no.~5, pp. 922--926, 2023.

\bibitem{9745078}
Y.~Wang, W.~Zhang, Y.~Chen, C.-X. Wang, and J.~Sun, ``Novel multiple
  {RIS}-assisted communications for 6g networks,'' \emph{{IEEE} Commun. Lett.},
  vol.~26, no.~6, pp. 1413--1417, 2022.

\bibitem{9868343}
Y.~Liu, L.~Zhang, F.~Gao, and M.~A. Imran, ``Intelligent reflecting surface
  networks with multiorder-reflection effect: System modeling and critical
  bounds,'' \emph{{IEEE} Trans. Commun.}, vol.~70, no.~10, pp. 6992--7005,
  2022.

\bibitem{mu2021simultaneously}
X.~Mu, Y.~Liu, L.~Guo, J.~Lin, and R.~Schober, ``Simultaneously transmitting
  and reflecting ({STAR}) {RIS} aided wireless communications,'' \emph{{IEEE}
  Trans. Wireless Commun.}, vol.~21, no.~5, pp. 3083--3098, May 2021.

\bibitem{9690478}
Y.~Liu, X.~Mu, J.~Xu, R.~Schober, Y.~Hao, H.~V. Poor, and L.~Hanzo, ``{STAR}:
  Simultaneous transmission and reflection for 360° coverage by intelligent
  surfaces,'' \emph{{IEEE} Wireless Commun.}, vol.~28, no.~6, pp. 102--109,
  2021.

\bibitem{10550177}
X.~Mu, J.~Xu, Z.~Wang, and N.~Al-Dhahir, ``Simultaneously transmitting and
  reflecting surfaces for ubiquitous next-generation multiple access in 6g and
  beyond,'' \emph{Proc. {IEEE}}, vol. 112, no.~9, pp. 1346--1371, 2024.

\bibitem{9863732}
J.~Zuo, Y.~Liu, Z.~Ding, L.~Song, and H.~V. Poor, ``Joint design for
  simultaneously transmitting and reflecting ({STAR}) {RIS} assisted {NOMA}
  systems,'' \emph{{IEEE} Trans. Wireless Commun.}, vol.~22, no.~1, pp.
  611--626, Jan. 2023.

\bibitem{wu2021coverage}
C.~Wu, Y.~Liu, X.~Mu, X.~Gu, and O.~A. Dobre, ``Coverage characterization of
  {STAR-RIS} networks: {NOMA} and {OMA},'' \emph{{IEEE} Commun. Lett.},
  vol.~25, no.~9, pp. 3036--3040, Sep. 2021.

\bibitem{9856598}
X.~Yue, J.~Xie, Y.~Liu, Z.~Han, R.~Liu, and Z.~Ding, ``Simultaneously
  transmitting and reflecting reconfigurable intelligent surface assisted
  {NOMA} networks,'' \emph{{IEEE} Trans. Wireless Commun.}, vol.~22, no.~1, pp.
  189--204, 2023.

\bibitem{10049110}
R.~Zhong, X.~Mu, Y.~Liu, Y.~Chen, J.~Zhang, and P.~Zhang, ``{STAR-RIS}s
  assisted {NOMA} networks: A distributed learning approach,'' \emph{{IEEE} J.
  Sel. Topics Signal Process.}, vol.~17, no.~1, pp. 264--278, 2023.

\bibitem{10050140}
Q.~Gao, Y.~Liu, X.~Mu, M.~Jia, D.~Li, and L.~Hanzo, ``Joint location and
  beamforming design for {STAR-RIS} assisted {NOMA} systems,'' \emph{{IEEE}
  Trans. Commun.}, vol.~71, no.~4, pp. 2532--2546, 2023.

\bibitem{9786807}
M.~Aldababsa, A.~Khaleel, and E.~Basar, ``{STAR-RIS-NOMA} networks: An error
  performance perspective,'' \emph{{IEEE} Commun. Lett.}, vol.~26, no.~8, pp.
  1784--1788, 2022.

\bibitem{9915477}
X.~Li, Y.~Zheng, M.~Zeng, Y.~Liu, and O.~A. Dobre, ``Enhancing secrecy
  performance for {STAR-RIS} {NOMA} networks,'' \emph{{IEEE} Trans. Veh.
  Technol.}, vol.~72, no.~2, pp. 2684--2688, 2023.

\bibitem{10049460}
F.~Karim, S.~K. Singh, K.~Singh, S.~Prakriya, and M.~F. Flanagan, ``On the
  performance of {STAR-RIS}-aided {NOMA} at finite blocklength,'' \emph{{IEEE}
  Wireless Commun. Lett.}, vol.~12, no.~5, pp. 868--872, 2023.

\bibitem{9838767}
Y.~Liu, X.~Mu, R.~Schober, and H.~V. Poor, ``Simultaneously transmitting and
  reflecting ({STAR})-{RIS}s: A coupled phase-shift model,'' in \emph{Proc.
  IEEE Int. Conf. Commun.}, Seoul, Republic of Korea, May 2022, pp. 2840--2845.

\bibitem{9849460}
J.~Zhao, Y.~Zhu, X.~Mu, K.~Cai, Y.~Liu, and L.~Hanzo, ``Simultaneously
  transmitting and reflecting reconfigurable intelligent surface ({STAR-RIS})
  assisted {UAV} communications,'' \emph{{IEEE} J. Sel. Areas Commun.},
  vol.~40, no.~10, pp. 3041--3056, 2022.

\bibitem{10050406}
Z.~Wang, X.~Mu, and Y.~Liu, ``{STARS} enabled integrated sensing and
  communications,'' \emph{{IEEE} Trans. Wireless Commun.}, vol.~22, no.~10, pp.
  6750--6765, 2023.

\bibitem{mei2021distributed}
W.~Mei and R.~Zhang, ``Distributed beam training for intelligent reflecting
  surface enabled multi-hop routing,'' \emph{{IEEE} Wireless Commun. Lett.},
  vol.~10, no.~11, pp. 2489--2493, 2021.

\bibitem{9723331}
E.~Bj{\"o}rnson, {\"O}.~T. Demir, and L.~Sanguinetti, ``A primer on near-field
  beamforming for arrays and reconfigurable intelligent surfaces,'' in
  \emph{Proc. 55th Asilomar Conf. Signals, Syst., Comput.}, 2021, pp. 105--112.

\bibitem{west2001introduction}
D.~B. West \emph{et~al.}, \emph{Introduction to graph theory}.\hskip 1em plus
  0.5em minus 0.4em\relax Prentice hall Upper Saddle River, 2001, vol.~2.

\bibitem{10086045}
W.~Mei and R.~Zhang, ``Joint base station and {IRS} deployment for enhancing
  network coverage: A graph-based modeling and optimization approach,''
  \emph{{IEEE} Trans. Wireless Commun.}, vol.~22, no.~11, pp. 8200--8213, 2023.

\bibitem{10439018}
M.~Fu, W.~Mei, and R.~Zhang, ``Multi-passive/active-{IRS} enhanced wireless
  coverage: Deployment optimization and cost-performance trade-off,''
  \emph{{IEEE} Trans. Wireless Commun.}, vol.~23, no.~8, pp. 9657--9671, 2024.

\end{thebibliography}

\end{document}